\documentclass[aps,prd,twocolumn,groupedaddress,showpacs]{revtex4}

\usepackage{graphics}

\begin{document}


\title{Mach's principle: 
Exact frame-dragging 
via gravitomagnetism in 
\\
perturbed Friedmann-Robertson-Walker universes 
with $K=(\pm 1, 0)$}



\author{Christoph Schmid}
\email{chschmid@itp.phys.ethz.ch}

\affiliation{ETH Zurich, Institute for Theoretical Physics, 
8093 Zurich, Switzerland}

\date{\today}

\begin{abstract}

We show that there is 
     {\it exact dragging}
of the axis directions of local
     {\it inertial frames} 
by a weighted average of the
cosmological energy currents 
via 
     {\it gravitomagnetism} 
for  
     {\it all} 
linear perturbations of 
     {\it all} 
Friedmann-Robertson-Walker (FRW) universes
and of Einstein's static closed universe, and for 
     {\it all}
energy-momentum-stress tensors and
in the presence of a cosmolgical constant.
This includes FRW universes arbitrarily close 
to the Milne Universe and the de Sitter universe. 
Hence the   
     {\it postulate formulated by Ernst Mach}
about the 
     {\it physical cause} 
for the time-evolution of inertial axes 
is shown to hold in General Relativity 
for linear perturbations of FRW universes.~---
The time-evolution of local inertial axes 
(relative to given local fiducial axes) 
is given experimentally
by the precession angular velocity  
$\vec{\Omega}_{\rm gyro}$ of local gyroscopes,
which in turn gives the operational definition of the 
gravitomagnetic field: 
$\vec{B}_{\rm g} \equiv -2 \, \vec{\Omega}_{\rm gyro}.$
The gravitomagnetic field is caused 
by energy currents 
$\vec{J}_{\varepsilon}$ via the 
       {\it momentum constraint,} 
Einstein's $G^{\hat{0}}_{\, \,  \hat{i}}$ equation,
$(-\Delta + \mu^2) \vec{A}_{\rm g} 
= - 16 \pi G_{\rm N} \vec{J}_{\varepsilon}$
with $\vec{B}_{\rm g} = \mbox{curl} \,  \vec{A}_{\rm g}.$
This equation 
is analogous to Amp\`ere's law, 
but it holds for all 
     {\it time-dependent} 
situations.
$\Delta$ is the 
     {\it de Rham-Hodge Laplacian,}
and $\Delta = - \mbox{curl} \, \mbox{curl}$ 
for the vorticity sector in Riemannian 3-space.~---
In the solution for an open universe
the $1/r^2$-force of Amp\`ere 
is replaced by a 
     {\it Yukawa force} 
$Y_{\mu} (r)  = (-d/dr)[(1/R) \exp (- \mu r)],$ 
     {\it form-identical}
for FRW backgrounds with $K = (-1, 0).$ 
Here $r$ is the measured geodesic distance 
from the gyroscope to the cosmological source, 
and $2 \pi R$ is the measured circumference of the sphere 
centered at the gyroscope and going through the source point. 
The scale of the exponential cutoff
is the $H$-dot radius, 
where $H$ is the Hubble rate, 
dot is the derivative with respect to cosmic time,
and $\mu^2 = -4 (dH/dt).$
Analogous results hold 
in closed FRW universes
and in Einstein's closed static universe.~---
We list six fundamental tests for the principle formulated by Mach:
all of them are explicitly fulfilled by our solutions.~---    
We show that only energy currents in the toroidal vorticity  sector 
with $\ell =1$ can affect the precession of gyroscopes. 
We show that the harmonic decomposition 
of toroidal vorticity fields 
in terms of vector spherical harmonics $\vec{X}^{-}_{\ell m}$
has radial functions which are    
        {\it form-identical} 
for the 3-sphere, the hyperbolic 3-space, and Euclidean 3-space, 
and are form-identical with the 
spherical Bessel-, Neumann-, and Hankel functions.~---  
The Appendix gives 
the de Rham-Hodge Laplacian 
on vorticity fields in Riemannian 3-spaces
by equations connecting 
the calculus of differential forms
with the curl notation.
We also give the derivation of the Weitzenb\"ock formula
for the difference between the  de Rham-Hodge Laplacian $\Delta$
and the ``rough'' Laplacian $\nabla^2$ on vector fields.  

\end{abstract}


\maketitle


\section{SUMMARY AND CONCLUSIOS}
\label{Introduction.Conclusions}


\subsection{Conclusions}
\label{Conclusions}


The time-evolution of 
       {\it local inertial axes,}
i.e. the local 
      {\it non-rotating frame},
is experimentally determined by the spin axes of 
     {\it gyroscopes,}
e.g. by inertial guidance systems. 
This has been first demonstrated by Leon Foucault in 1852.


It is an 
    {\it observational fact} 
that the spin axes of gyroscopes 
far away from Earth 
do not precess relative to quasars.
This observational fact has been named 
    {\it ``Mach zero''}
by Bondi and Samuel
    \cite{Bondi.Samuel}.~--- 
Near Earth there are two corrections predicted by General Relativity: 
(1) The de Sitter precession, which is due to the gyroscope's motion in
      the curved Schwarzschild metric, and which has been measured.
(2) The extremely small frame-dragging effect by the rotating Earth,
      the Lense-Thirring effect,
      which is predicted to be 43 milli-arc-sec per year,
      which 
      hopefully will be extracted from the data taken 
      by Gravity Probe B 
                 \cite{GP.B},
      and whose measurement by the LAGEOS satellites 
      has been reported in
          \cite{LAGEOS}.~---  
The dragging effect by the rotating Earth 
(and also the perihelion shift of Mercury) 
is 
    {\it measured relative to distant quasars.} 
Therefore a measurement by Gravity Probe B 
and the LAGEOS satellites
(and a measurement of the perihelion shift)  
is a  
     {\it test of two things combined,}
on the one hand a test of Einstein's General Relativity, 
on the other hand a test of the observational fact ``Mach zero''.


But the fundamental question
remains: 
   {\it What physical cause}
explains 
the observed time-evolution of gyroscope axes,
the observational fact ``Mach zero''? 
In J.A. Wheeler's words: 
Who gives the marching orders
to the spin axes of gyroscopes, i.e. to inertial axes?
Can theory predict resp. explain the extremely precise
observational tests of ``Mach zero''?


An answer to this fundamental question was   
    {\it formulated by Ernst Mach} 
   \cite{Mach.Mechanik, Mach.Energy}
in his  
   {\it postulate} 
that inertial axes 
   {\it exactly follow}
an average of the motion of masses in the Universe.
Mach's postulate demands 
   {\it exact frame dragging,}
not merely a little bit of frame dragging as in the
Lense-Thirring effect.~---
Mach did not have a mechanism for dragging.
The new force came with General Relativity: 
Gravitomagnetism, exemplified in the Lense-Thirring effect.~--- 
Mach also did not know, 
what average
of the cosmological mass currents should be taken.


Mach's starting postulate was:
    {\it No absolute space}, 
in modern parlance
    {\it no absolute element}
in the input.
Only relative motion of physical objects is significant.~---  
From this requirement follows 
Mach's postulate of
    {\it frame invariance} 
of observable predictions
under time-dependent rotations
of the reference frame.
In Mach's words
   \cite{Mach.Energy}:
\begin{quote}
  ``Obviously it does not matter, 
    whether we think of the Earth rotating around its axis,
    or if we imagine a static Earth
    with the celestial bodies rotating around it.'' 
\end{quote}
From these requirements follows 
Mach's postulate 
    \cite{Mach.Mechanik}:
      Centrifugal forces and 
      the rotation of the plane of Foucault's pendulum
      are 
         {\it totally determined} 
    by the relative motion 
    of masses in the universe  
    with respect to 
Earth-fixed axes.
In modern parlance:
    The time-evolution of local inertial axes is 
       totally determined
    by the energy currents in the universe including
    the effect of gravitational waves.~---
The demands ``no absolute element,'' 
            ``frame invariance under rotational motion,'' and
            ``totally determined''
necessitate the postulate of
      {\it exact dragging}
of inertial axes by energy currents in the universe. 
With only partial dragging it would be impossible to fulfill
any of the three postulates "no absolute element in the input",
"frame invariance of the solutions", and
"time evolution of inertial axes totally determined
by cosmological energy currents".

For a purely
   {\it local analysis,}
if one has only laboratory experiments available
(like Newton's bucket experiment),
the local nonrotating frame is
an absolute element.~---  
But, in contrast, for a
   {\it global analysis,}
if one considers observations of the entire universe,
the local nonrotating frame 
is not an absolute element in the input,
because (according to Mach)
the nonrotating frame is totally determined
by mass motions (energy currents) in the universe,
it is nonrotating
   {\it relative}
to an average of the energy currents
in the universe.

The
       {\it ``relativists''}
(Huyghens, Berkeley, Leibniz, Mach, and others)
wanted  a theory
       without absolute space
(without an absolute element in the input),
a theory, where only relative motion is significant.
After Einstein's theory of General Relativity
it was soon recognized that this theory,
when applied to the solar system, is
not a ``Theory of Relativity''
in the sense that the solutions violate the postulates
``no absolute element'',
``solutions frame invariant'',
``totally determined'', and
``exact dragging'',
see Sec.~\ref{sect.frame.invariance}.
This is why Einstein and others started
to focus on cosmological models.
The unperturbed Friedmann-Robertson-Walker models
satisfy Mach's postulates.
In this paper we derive the new result
that all FRW models with arbitrary linear perturbations
also satisfy Mach's postulates.
This means that
      {\it Cosmological General Relativity}
(at the level of linear perturbations of FRW universes)
is a true
      {\it "Theory of Relativity".}


We have shown in
         \cite{Rovaniemi, CS.I} 
that the axis directions of local 
         {\it inertial frames}         
everywhere in the universe  
         {\it exactly follow}  
a weighted average of the  
energy currents in the universe,
i.e. there is 
       {\it exact frame-dragging,}          
for linear perturbations  
of a Friedmann-Robertson-Walker (FRW) background 
with $K=0$ (i.e. spatially flat).
The 
        {\it weight function} 
has an exponential cutoff
at the $H$-dot radius, 
where $H$ is the Hubble rate, 
and dot is the derivative with respect to measured cosmic time. 
Hence we have demonstrated 
the validity of the hypothesis formulated by Mach 
about the 
        {\it physical cause} 
for the time-evolution 
of inertial-frame axes  
for linear perturbations of a FRW background with $K=0.$
The proof follows from Einstein's $G_{\hat{0} \hat{i}}$ equation, 
the 
       {\it momentum constraint.} 
The dynamical mechanism is 
       {\it cosmological gravitomagnetism.}


In the present paper we show 
that results analogous to those derived for $K=0$ in 
        \cite{CS.I}  
are valid for linear perturbations of 
FRW backgrounds with $K= \pm 1,$ 
i.e. backgrounds which are 
spatially spherical ($S^3$) resp. hyperbolic
($H^3$).


Many alternative and inequivalent postulates have been proposed 
under the heading ``Mach's principle'' by many authors
after E. Mach. 
An incomplete list of 10 inequivalent versions of 
Mach's principle  
has been given by Bondi and Samuel
   \cite{Bondi.Samuel}
(based in part on  
   \cite{Barbour.Pfister}).
While the versions ``Mach~9'' and ``Mach~10''
listed by 
   \cite{Bondi.Samuel} 
are closely related to our ``Fundamental Tests~3~and~4'' in  
   Sec.~\ref{fundamental.tests},
their versions 
   ``Mach~1-8'' 
definitely do not correspond 
to anything in the writings of Mach.
Furthermore 
each one of their versions
   ``Mach~1-8'' is 
either not true in cosmological General Relativity, 
as stated already in
    \cite{Bondi.Samuel},
or irrelevant, as we explain in 
   Sec.~\ref{inequivalent.postulates}.


A more detailed introduction to the conceptual issues 
of the postulate formulated by Mach
and discussions of publications  
     \cite{Thirring}-\cite{Bicak}
(mostly cosmological as postulated by Mach)
are given in Secs.~I and II of the companion paper
     \cite{CS.I}.
Additional references are given in
\cite{Barbour.Pfister, C.Klein, Lynden-Bell.1995, Ciufolini.Wheeler, 
Pfister.history}.     
Comparisons of our methods and results with 
the recent papers by 
Bi\v{c}\'ak, Lynden-Bell, and Katz
     \cite{Bicak, Bicak.2007}
are at the end of 
   Sects.~\ref{exponential.cutoff} and
          \ref{fundamental.tests}.~---  
We use the conventions of Misner, Thorne, and Wheeler
   \cite{MTW}.


\subsection{Vorticity perturbations}
\label{Vorticity.Perturbations}


We now summarize the main results of the present paper
for the case when only vorticity perturbations
(= 3-vector perturbations, by definition divergenceless) 
are present.


Two important results  
from cosmological perturbation theory
    \cite{Bardeen} 
for $K = (0, \pm 1)$
are needed 
to understand the following summary.
In the 
    {\it vorticity sector:}
\begin{enumerate}
\item The 
     {\it slicing} 
   of space-time in slices $\Sigma_t$ of fixed time,   
   i.e. $3$-spaces, is 
     {\it unique}. 
   The 
     {\it lapse} 
   function (elapsed measured time between slices) and $g_{00}$ are 
     {\it unperturbed}. 
\item The 
     {\it intrinsic geometry} 
   of each slice $\Sigma_t,$ 
   i.e. of 3-space, remains
     {\it unperturbed.}
\end{enumerate}
Details and proofs are given in Sec. III of
      \cite{CS.I}.
Note that Secs. III - V of
      \cite{CS.I}
are written for FRW backgrounds with $K = (0, \pm 1).$


For the 3-geometry, 
which remains unperturbed ($E^3,$ resp. $S^3,$ resp. $H^3$),
and for our fixed-time problem (the momentum constraint),
the uniquely appropriate coordinate choice (gauge choice)
is comoving spherical coordinates,
$x^i = (\chi, \theta, \phi),$
where $(a \chi)$ is the measured radial distance
with $a(t)=$ scale factor.
With this gauge choice the coefficients of the
      {\it 3-metric} 
$\, ^{(3)}g_{ij}$ are 
      {\it also unperturbed.} 
The spatial coordinate system is 
      {\it geometrically rigid} 
apart from the uniform Hubble expansion. 
Also $g_{00}$ is unperturbed, and we choose the time-coordinate $t$
to be the measured cosmic time.
Hence, with our gauge choice the 
   {\it only} 
quantity referring to vorticity perturbations is
the 
   {\it shift} 
3-vector $\beta^i$ (resp $\beta_i = g_{0i}$),
\begin{eqnarray}
ds^2 &=& - dt^2 + a^2 
[\, d\chi^2 + R^2_{\rm com} ( d\theta^2 + \sin^2\theta \, d\phi^2  ) ]
\nonumber
\\
&& + 2 \beta_i \, dx^i dt,
\label{metric}
\end{eqnarray}
where $R_{\rm com}$ stands for $R_{\rm comoving},$ and
$R_{\rm com} (\chi) = (\chi, \, \sin \chi, \, \sinh \chi)$
for $K = (0, +1, -1).$ 
The index of the shift 3-vector is lowered and raised 
by the 3-metric, e.g. $\beta_i = g_{ij} \beta^j.$


Our choice of unperturbed comoving 
spherical coordinates for $(E^3, S^3, H^3)$
is 
     {\it singled out} 
by the fact that 
our radial coordinate lines ($\theta$ and $\phi$ fixed) are 
     {\it geodesics} 
in 3-space, an important property in discussions of Mach's principle. 
In all other gauges,
where perturbed 3-metrics $g_{\mu \nu}$ are used,
the radial coordinate lines 
wind up more and more
into ``spirals'' (as time goes on)
relative to geodesics on $\Sigma_t$.
Hence all other gauges (particularly time-orthogonal) 
give a very awkward way 
to coordinatize unperturbed $(E^3, S^3, H^3)$ spaces.


Up to now the coordinatization (gauge) is not yet completely fixed,
there still exists the freedom of 
     {\it residual gauge transformations} 
by time-dependent, spatially 
     {\it rigid rotations}  
of the coordinates for 3-space.
Rigid rotations are crucial in Mach's statement on frame-invariance
   \cite{Mach.Energy}:
See the quote in Sec.~IA, "Obviously it does not matter ...".

We shall show in
     Sec.~\ref{sect.frame.invariance}
that our solution 
of the equations of 
cosmological gravitomagnetism is explicitly 
    {\it form-invariant}
under time-dependent, spatially rigid rotations 
of the reference frame,
i.e. 
    {\it frame-independent}
(valid in all reference frames
connected by time-dependent rigid rotations).
In contrast, the 
    {\it solutions} 
of the equations of General Relativity for the 
    {\it solar system}
are 
    {\it not} 
form-invariant under spatially rigid, time-dependent rotations.


Measuring the angular velocity of a given celestial body
requires that one first 
fixes the residual gauge freedom, 
i.e. one must first state, 
relative to what one wants to measure the angular velocity.
This can be done by making a definite choice 
of spatial axis directions along 
   {\it one world line.}
In the spirit of Mach 
one could fix the axis directions
either along the world-line of the Earth to Earth-fixed axes,
or   (in an open, asymptotically unperturbed universe)
to asymptotic quasars.


\subsection{Gravitomagnetic and gravitoelectric fields}
\label{gravitomagn.fields}


The 
   {\it general operational definitions}
(valid beyond perturbation theory)
of the gravitomagnetic field
$\vec{B}_{\rm g}$ and the gravitoelectric field
$\vec{E}_{\rm g}$
are given
via measurements by  
    {\it fiducial observers} 
(FIDOs)
with nonintersecting world lines
and with their local ortho-normal bases, LONBs
    \cite{Thorne}. 
These FIDOs measure 
the first time-derivatives of the LONB components 
of the 3-momentum, $p_{\hat{i}},$ 
for free-falling quasistatic test particles, resp.
of the spin-vector, $S_{\hat{i}},$ 
for comoving gyroscopes, 
\begin{eqnarray} 
\mbox{free-falling quasistatic test particle:} 
\, \, \, \, \frac{d}{dt} \, (p_{\hat{i}}) 
\equiv m E_{\hat{i}}^{\rm g}, &&
\label{def.E}
\\
\mbox{gyro comoving with FIDO:} 
\, \, \, \, \frac{d}{dt} (S_{\hat{i}}) 
\equiv
- \frac{1}{2} (\vec{B}_{\rm g} \times \vec{S})_{\hat{i}}, &&
\label{op.def.B} 
\\
\Omega_{\hat{i}}^{\rm gyro}  
= - \frac{1}{2}  B_{\hat{i}}^{\rm g}, &&
\label{def.B} 
\end{eqnarray}
where $\Omega_{\hat{i}}$ are the LONB components of
the precession rate of the spin axes of a gyroscope.
Hats over indices denote components in a
local ortho-normal basis (LONB),  
and $t$ is the local time measured by the FIDO.
Note that LONB components, e.g. $p_{\hat{i}},$
are 
     {\it directly measurable} 
(for a given field of LONBs), 
while coordinate-basis components, e.g. $p_{\mu},$
are 
  {\it not measurable without prior knowledge of the metric} 
$g_{\mu \nu}.$


The operational definitions of 
$\vec{B}_{\rm g}$ and $\vec{E}_{\rm g}$
are 
      {\it identical} 
with the operational definitions 
of the ordinary electric and magnetic fields
via measurements on charged point test-particles and
on charged spinning test particles, 
except that the charge $q$ is replaced by the mass $m$ for
quasistatic test particles.


The $(\vec{B}_{\rm g}, \vec{E}_{\rm g} )$-fields
depend on the choice of FIDOs.
For free-falling FIDOs 
$\vec{E}_{\rm g} \equiv 0,$ 
and for nonrotating FIDOs (relative to axes of local gyroscopes)
$\vec{B}_{\rm g} \equiv 0,$ 
which are the twin conditions in the equivalence principle.~---  
The $( \vec{E}_{\rm g}, \vec{B}_{\rm g} )$ fields
are directly related to 
connection coefficients:
$
(\omega_{\hat{i}\hat{0}})_{\hat{0}} \equiv
- E_{\hat{i}}^{\rm g}, \quad 
(\omega_{\hat{i}\hat{j}})_{\hat{0}} \equiv
- \frac{1}{2}  B_{\hat{i}\hat{j}}^{\rm g},
\label{2nd.op.def.EB}
$
where $B_{\hat{i}\hat{j}} \equiv 
\varepsilon_{\hat{i}\hat{j}\hat{k}}B_{\hat{k}}, \,
\Omega_{\hat{i}\hat{j}} \equiv 
\varepsilon_{\hat{i}\hat{j}\hat{k}} \Omega_{\hat{k}},$
and
$\Omega_{\hat{i}\hat{j}} = (\omega_{\hat{i}\hat{j}})_{\hat{0}}.$
More details about 
the general operational definitions of 
$( \vec{E}_{\rm g}, \vec{B}_{\rm g} )$
are given in Sec.~IV of
      \cite{CS.I}.


Our 
    {\it specific choice}
for a field of 
    {\it FIDOs:}
Our FIDOs are at fixed values of our coordinates $x^i,$
and we fix the spatial axis directions of our FIDOs 
in the directions of our coordinate basis vectors
$(\partial / \partial \chi, \,
  \partial / \partial \theta, \,  
  \partial / \partial \phi).$~--- 
We recall from 
    Sec.~\ref{Vorticity.Perturbations}
that our choice of unperturbed comoving 
spherical coordinates for $(E^3, S^3, H^3)$
is 
     {\it singled out} 
by the fact that 
the radial coordinate lines ($\theta$ and $\phi$ fixed) are 
     {\it geodesics} 
in 3-space. 
Additional details about our specific choice of FIDOs  
are given in 
      \cite{CS.I}
at the end of Sec.~IV.


The 
      {\it gravitomagnetic vector potential} 
$\vec{A}_{\rm g}$ 
in the vorticity sector is 
     uniquely determined 
by $\vec{B}_{\rm g} \equiv {\rm curl} \, \vec{A}_{\rm g},$ 
because div~$\vec{A}_{\rm g} \equiv 0.$~---  
With $\beta_i \equiv g_{0i}$ and $A_i^{\rm g}$ 
defined independently above,
it follows that the 
     {\it shift vector} 
$\vec{\beta}$ 
is equal to the 
     {\it gravitomagnetic vector potential} 
$\vec{A}_{\rm g},$
see also Sec. V of
    \cite{CS.I}.~---   Our gravitomagnetic vector potential 
$\vec{A}_{\rm g}$
is directly proportional to 
    {\it Bardeen's gauge-invariant amplitude} $\Psi$ 
as shown in 
     Eq.~(\ref{our.shift.Psi}).


In 
     Sec.~\ref{Riemann.Efield.Bfield}
we shall consider gravitomagnetism on a Minkowski background 
and show that our FIDOs (and hence our 
$\vec{E}_{\rm g}$ and $ \vec{B}_{\rm g}$ fields) 
are  
     {\it singled out} 
by the property that 
all 13 nonzero components of the 
     {\it Riemann tensor} 
are 
     {\it uniquely determined}  
by spatial and time derivatives 
of our $\vec{B}_{\rm g}$ field alone.
Our $\vec{B}_{\rm g}$ field 
gives an efficient and physically transparent way 
to represent all information in the Riemann tensor.~---  
In the opposite direction: 
We shall show that our  $\vec{B}_{\rm g}$ field 
is determined by the
Ricci components $R_{\hat{k}\hat{0}}$ 
up to a homogeneous $\vec{B}_{\rm g}$ field,
which is equivalent to a 
time-dependent rigid rotation of the coordinates and FIDO axes
according to
   Eq.~(\ref{def.B}).


\subsection{The momentum constraint}


The crucial equation for Mach's principle is the 
momentum constraint, 
Einstein's $G^{\hat{0}}_{\, \, \hat{i}}$ equation,
for cosmological vorticity perturbations.
The momentum constraint 
for perturbations of
    {\it all three} 
FRW background geometries, $K = (\pm 1, 0),$
is given by
    {\it one form-identical equation without curvature terms:} 
\begin{equation}
(- \Delta + \mu^2) \,   \vec{A}_{\rm g} 
= - 16 \pi G_{\rm N} \, \vec{J}_{\varepsilon}, 
\label{Einstein.equation.A.introduction}
\end{equation}
where 
$(\mu/2)^2 \equiv - (dH/dt) \equiv (\mbox{Hubble-dot radius})^{-2}.$
Derivation of 
     Eq.~(\ref{Einstein.equation.A.introduction}) 
in
     Sec.~\ref{momentum.constraint}.
Our momentum constraint
has the form of the 
     {\it inhomogeneous Helmholtz equation}  
with the sign of the $\mu^2$-term opposite to the usual sign.
The sign in
    Eq.~(\ref{Einstein.equation.A.introduction}) 
causes an exponential cutoff, which becomes 
important beyound the Hubble-dot radius.~---    
The source $\vec{J}_{\varepsilon}$ with
$ J_{\hat{i}}^{\varepsilon}  \equiv 
T^{\hat{0}}_{\, \, \, \hat{i}}$ 
for all types of matter is the directly  
    {\it measurable} energy-current 
density,
which is equal to the  
measurable momentum
density (for $c = 1$), 
hence the name ``momentum constraint''.
``Directly measurable'' means 
that 
$J_{\hat{i}}^{\varepsilon}$
is measurable 
    {\it without prior knowledge} 
of $g_{0i},$ which is the output of solving 
the momentum constraint.~---  
The energy current  $\vec{J}_{\varepsilon}$ 
is analogous to the 
charge current $\vec{J}_q$ in Amp\`ere's law.
For perfect fluids and linear perturbations
$\vec{J}_{\varepsilon} = (\rho + p) \vec{v}.$~---  
All symbols in
    Eq.~(\ref{Einstein.equation.A.introduction})
refer to the physical scale, they are not rescaled to
a comoving length scale.
Therefore the scale factor 
$a(t)$ does 
        {\it not} 
appear in 
        Eq.~(\ref{Einstein.equation.A.introduction}).


It is remarkable that 
   Eq.~(\ref{Einstein.equation.A.introduction})
holds for 
     {\it time-dependent} 
gravitomagnetodynamics.
All the same, it is an 
    {\it elliptic} partial differential equation: 
{\it No partial time-derivatives of the perturbation field} 
$\vec{A}_{\rm g}.$
Time derivatives only appear in the given background, 
$H=a^{-1}  da/dt, \, \dot{H} = dH/dt.$~---  
     Eq.~(\ref{Einstein.equation.A.introduction}) 
has the same form as Amp\`ere's law
for stationary magnetism
except for the $\mu^2 \vec{A}_{\rm g}$ term.~---   
The analogue of the Maxwell term $(\partial_t \vec{E})$ 
in the Amp\`ere-Maxwell equation is 
    {\it absent} 
for the 
    {\it time-dependent}  
context of gravitomagnetism.
No inconsistency arises when taking the divergence of 
    Eq.~(\ref{Einstein.equation.A.introduction}),
since  
div~$\vec{A}_{\rm g} \equiv 0$ 
and   div~$\vec{J}_{\varepsilon} \equiv 0$
in the vorticity sector.
The analogue of a Maxwell term 
cannot be present in gravitomagnetism,  
this would produce gravitational vector waves, which is impossible.
Additional explanations about the momentum constraint 
are given after Eq.~(36)
in Sec.~VI of
    \cite{CS.I}.
 

Other perturbed Einstein equations 
and the perturbed evolution equations for matter 
will not be needed in the context of this paper.
The full set of field equations for 
linear cosmological gravitomagnetism
is given in 
   Secs. V and VI of 
    \cite{CS.I}.


\subsection{The de Rham-Hodge Laplacian on vector fields 
in Riemannian 3-spaces}
\label{subsection.Laplacian}


The 
     {\it absence} 
of a 
     {\it 3-curvature term} 
in
     Eq.~(\ref{Einstein.equation.A.introduction})
is connected to our use of the 
     {\it de Rham-Hodge Laplace} operator $\Delta,$
which is called ``the'' Laplace operator in the 
literature on differential forms in Riemannian geometry
    \cite{Jost}-\cite{de.Rham}.
Except when acting on scalar fields,
the de Rham-Hodge Laplace operator $\Delta$
must be distinguished from 
$\nabla^2 = g^{\mu \nu} \nabla_{\mu} \nabla_{\nu},$  
named the ``rough Laplacian'' in
   \cite{Frankel, Berger}.~---
Unfortunately $\nabla^2$ has been used 
in all publications  
on cosmological vector perturbations 
(as far as we know),
e.g. in 
     \cite{Bardeen, Tomita},
and moreover $\nabla^2$ has been  
called Laplacian and/or denoted by $\Delta$
in many publications on vector perturbations,         
e.g. in
      \cite{Kodama.Sazaki}-\cite{Hu.Seljak.White.Zaldarriaga}.
Using $\nabla^2$ has the unfortunate consequence that
curvature terms appear, where none are present 
when the de Rham-Hodge Laplacian $\Delta$ is used.
The most important example is electromagnetism 
in curved space-time,
where the equivalence principle forbids curvature terms,
if the de Rham-Hodge Laplacian 
resp d'Alembertian is used
     \cite{MTW.deRham}.


In Riemannian 3-space, the de Rham-Hodge Laplacian  
acting on vorticity fields,
i.e. fields with 
div~$\vec{a} 
= \star \, d \star \tilde{a} = \delta \tilde{a}
= 0,$ 
is given by
\begin{equation}
(\Delta \, \vec{a})_{\mu} 
= - ({\rm curl \, curl} \, \vec{a})_{\mu}
= - (\star \, d \star d \, \tilde{a})_{\mu},
\label{Laplace.introduction}
\end{equation}
where we have given 
both the notation of elementary vector calculus
and  the notation of differential forms 
with $d \equiv $ exterior derivative,
$(-\delta) \equiv d^* \equiv$ adjoint operator to the exterior derivative, 
$\star \equiv $ Hodge dual, 
tilde $\tilde{} \,$ denoting $p$-forms,
and $(\vec{a})_{\mu} \equiv (\tilde{a})_{\mu}.$
Note that 
$( \mbox{curl} \, \vec{a} )_{\mu} = (\star \, d \, \tilde{a})_{\mu}.$
Details are given in the
      Appendix~\ref{appendix.diff.forms}.

The de Rham-Hodge Laplacian is 
     {\it singled out} 
among the elliptic differential operators 
by the following properties:
(1) If in some region all types of sources are zero,
 div~$\vec{v} = 0 $ and           
curl~$\vec{v} = 0,$
then also   $\Delta~\vec{v} = 0.$
(2) $\Delta$ commutes with $d, \delta$ and~$\star.$
Therefore 
applying curl, div, and grad
to an eigenfield of  the Laplacian produces again 
an eigenfield of the
Laplacian with the same eigenvalue.~---
The ``rough Laplacian''~$\nabla^2$
does not have properties (1) and (2), 
if the Ricci tensor is not zero.~---
The de Rham-Hodge Laplacian is also easier to compute 
than the ``rough Laplacian'',
because no Christoffel symbols resp. connection coefficients 
are needed to compute it.


The difference between the de Rham-Hodge Laplacian and the 
``rough Laplacian'' 
is given by the 
     {\it Weitzenb\"ock} 
formula,
i.e. by curvature terms.
The Weitzenb\"ock formula for vector fields, 
is derived in
    Sec.~\ref{section.Weitzenboeck}
of the Appendix. 
For a FRW universe with $K = \pm 1 $ and
with the curvature radius $a_{\rm c} \equiv 1,$
the Weitzenb\"ock formula gives
$\Delta \vec{A} = (\nabla^2 - 2 K) \vec{A}.$


The de Rham-Hodge Laplacian applies to 
totally antisymmetric tensor fields.
For 
    {\it totally symmetric} 
tensor fields, e.g. for cosmological tensor perturbations, 
the appropriate Laplacian is the
    {\it Lichnerowicz Laplacian}
    \cite{Lichnerowicz.Laplacian, Lichnerowicz.LesHouches}.
For vector fields 
the Laplacians of Lichnerowicz and of de Rham-Hodge
coincide.


\subsection{No absolute element in the input: 
\\
The exponential cutoff at the $H$-dot radius}
\label{exponential.cutoff}


The 
    {\it crucial difference} 
between our momentum constraint,
     Eq.~(\ref{Einstein.equation.A.introduction}),
and the corresponding equation of 
Bardeen,
     Eq.~(4.12),
and the equation advocated by
Bi\v{c}\'ak et al
    \cite{Bicak, Bicak.2007}
is our $ \mu^2 \vec{A}_{\rm g}$ term, 
which causes the 
     {\it Yukawa cutoff}
in
     Eq.~(\ref{B.general.introduction}) 
below, and which is absent in
the solution of the equation advocated by
by Bi\v{c}\'ak  et al.
Our $\mu^2 \vec{A}_{\rm g}$ term is due to the difference of the 
     {\it sources}.


Our guiding principle and starting point: 
   {\it ``No Absolute Element in the Input''.}
The source, the matter 
     {\it input}, 
must be
     {\it measurable without prior knowledge} 
of $g_{0i} = A^{(\rm g)}_i .$ 
The latter is the
     {\it output}
of solving the momentum constraint.
From 
   Eq.~(\ref{def.B}) we see that
$\vec{B}_{\rm g} = {\mbox {curl}}  A_{\rm g}$ 
directly gives the local nonrotating frame all over the universe,
which should not be in the input, 
because it would be an absolute element in the input.~---
We write the momentum constraint
in terms of the  
    {\it LONB components}  $T^{\hat{0}}_{\, \,  \hat{i}},$
which are measurable 
without prior knowledge of $\vec{A}_{\rm g},$ 
i.e. without an absolute element in the input.


For Mach's principle, 
the relevant factor in the source is the 
     {\it angular velocity} 
of cosmic matter, 
    $\vec{\Omega}_{\rm matter},$ 
measured from our position and 
   {\it relative to geodesics} (on $\Sigma_t$)
which start in the directions of our chosen fiducial axes.
In the
    {\it local ortho-normal bases}, 
the transverse part of the energy current 
$T^{\hat{0}}_{\, \,  \hat{i}} = J_{\hat{i}}^{(\varepsilon)}$
is directly proportional to 
$\Omega_{\hat{i}}^{\rm matter}$ (relative to our fiducial axes), 
which can be measured 
     {\it without prior knowledge} 
of $g_{0i} = A^{(\rm g)}_i.$
The measured transverse matter velocity is  
directly proportional to Bardeen's 
     {\it gauge-invariant} 
velocity amplitude $v_{\rm s},$ 
which he does {\it not} use in his momentum constraint.


In contrast, Bardeen's
     Eq.~(4.12)
(and the equation advocated by
Bi\v{c}\'ak et al)
is written in the 
     {\it coordinate basis} 
with the source $T^{\, 0}_{\, \, i},$ 
and this is proportional to Bardeen's 
     {\it other} 
gauge-invariant velocity amplitude $v_{\rm c}.$
The curl of the velocity field
with the amplitude $v_{\rm c}$
is proportional to the angular velocity of matter 
    {\it relative} 
to the 
    {\it axes of gyroscopes} 
at the positions of the cosmic sources. 
But this has the severe drawback that $v_{\rm c}$ is 
    {\it not measurable} 
    {\it without prior knowledge} 
of the time-evolution of gyroscope axes all over the cosmos, 
which should be the 
   {\it output}
of solving the momentum constraint:  
$\vec{\Omega}_{\rm gyro} =  - \frac{1}{2} \vec{B}_{\rm g} 
=  - \frac{1}{2} {\mbox{curl}} \vec{A}_{\rm g}$
with
$A^{(\rm g)}_i = g_{0i}.$
Bardeen's version of the momentum constraint 
(and the version advocated by
Bi\v{c}\'ak et al)
needs an 
    {\it absolute element in the input}:
the local non-rotating frames all over the cosmos.
This violates the crucial starting demand:
No absolute element in the input.
The requirement  ``the theory should not contain an absolute element'', 
discussed in 
   \cite{Barbour.Pfister.Ehlers},
is closely connected to Mach's starting point: {\it No absolute space}.


Einstein commented about the energy-momentum tensor $T_{\mu \nu}$
in a 
      {\it coordinate basis:} 
``If you have a tensor $T_{\mu \nu}$
and not a metric ... 
the statement that matter by itself determines the metric
is meaningless''
     \cite{Einstein.matter.metric}.
This objection of Einstein directly applies to the approach advocated by
    Bi\v{c}\'ak  et al
    \cite{Bicak, Bicak.2007}.
On the other hand we have shown 
in Sec.~IX of
    \cite{CS.I}
that this objection of Einstein does not apply to the 
    {\it LONB components} 
of the energy-momentum tensor.


\subsection{The solution of the momentum constraint}
\label{introduction.solution}


The solution of the momentum constraint 
     Eq.~(\ref{Einstein.equation.A.introduction})
for cosmological gravitomagnetism on a background of
        {\it open} 
FRW universes
is given by 
        {\it identical expressions}
for  $K= 0$ and $K= -1,$
\begin{eqnarray}
\vec{B}_{{\rm g}} (P) 
=-4G_{\rm N} \int d({\rm vol}_Q)  
[\vec{n}_{PQ} \times \vec{J}_{\varepsilon}(Q)] 
 Y_{\mu} (r_{PQ}),
\label{B.general.introduction}
\\
Y_{\mu} (r) 
= 
\frac{-d}{dr} [\frac{1}{R} \exp (- \mu r)] 
= {\rm Yukawa \, \, force},
\label{Yukawa.force.introduction}
\end{eqnarray}
where $r_{PQ}$ is the 
measured geodesic (radial) distance 
from the field  point $P$ (gyroscope)
to the source point $Q,$ 
and $(2 \pi R_{PQ})$ is the 
measured circumference
of the great circle through $Q$ with the center $P.$
Hence $r=a \chi,$ where $a$ is the scale factor,
and $\chi$ is the radial coordinate 
(comoving geodesic distance from the origin).~---
The structure of 
   Eq.~(\ref{B.general.introduction})
is analogous to  
      {\it Amp\`ere's force} 
for ordinary stationary magnetism,
however the  $1/r^2$ of Amp\`ere's force
is replaced by the 
      {\it Yukawa force} of
   Eq.~(\ref{Yukawa.force.introduction}).~---
In the curved space $H^3$
the geodesic distance $r_{PQ}$ 
appears in the exponential cutoff 
and in the derivative of
     Eq.~(\ref{Yukawa.force.introduction}),
while $R_{PQ}$ appears in the denominator.


Although the structure of 
   Eq.~(\ref{B.general.introduction})
is analogous to the solution of 
Amp\`ere's law 
for ordinary stationary magnetism,
there is a 
   {\it fundamental difference}
to Amp\`ere's law,
since    
  Eq.~(\ref{B.general.introduction})
is valid for 
   {\it time-dependent} 
situations, i.e. gravitomagneto-dynamics.


The 
    {\it vector structure} 
of
    Eq.~(\ref{B.general.introduction}) 
needs explanation,
since there is no global parallelism 
on a hyperbolic 3-space $H^3.$  
The vectors $\vec{J}_{\varepsilon}(Q)$
and         $\vec{n}_{PQ}$
are in the tangent space at the source point $Q,$
and $\vec{n}_{PQ}$ is the unit vector pointing 
along the geodesic from $P$ to $Q.$
We parallel transport 
$[\vec{n}_{PQ} \, \times \, \vec{J}_{\varepsilon}(Q)]$
along the geodesic from $Q$ to $P,$ 
where the contributions from all sources are added.~---
We use the arrow-notation of Thorne et al
    \cite{Thorne}
for vectors in tangent spaces to Riemannian 3-spaces 
from a $(3+1)$-split.
This emphasizes 
the structure analogous to Amp\`ere's law.


\subsection{Frame invariance, no boundary conditions}
\label{sect.frame.invariance}


There is a second  
    {\it fundamental difference} 
between the solution 
   Eq.~(\ref{B.general.introduction})
for cosmological gravitomagnetism 
and the corresponding solutions in other theories,
Amp\`ere's magnetism, 
electromagnetism in Minkowski space,
and general relativity for the solar system.
The 
     {\it solutions} 
in all these 
     {\it other} 
theories contain an 
     {\it absolute element}:
Amp\`ere's law and 
the laws of electromagnetism in Minkowski space 
do not hold in frames which are rotating 
relative to inertial frames,
unless one introduces fictitious forces.


Einstein's 
    {\it equations} 
by themselves are form-invariant.
But for general relativity of the solar system,
when working in rotating frames  
relative to the asymptotic nonrotating frame
(the absolute element),
one must impose
    {\it boundary conditions} 
on the solutions in the asymptotic Minkowski space.
The boundary conditions  
    {\it encode the absolute element,} 
they explicitely encode fictitious forces 
(which do not have sources within the solar system),
as explained in Sec. XI of
    \cite{CS.I}.
Hence the 
     {\it solution} 
of Einstein's equations for the solar system is 
     {\it not form-invariant} 
when going to a frame 
which is rotating relative to asymptotic inertial frames.


In contrast, the solution for 
     {\it cosmological} 
General Relativity,
     Eq.~(\ref{B.general.introduction}), 
remains valid as it stands, i.e. the 
   {\it solution is form-invariant,}
if one goes to a frame which is in globally rigid rotation 
relative to the previous reference frame:
The form of the solution is 
      {\it frame-independent}, 
because both sides of 
    Eq.~(\ref{B.general.introduction})
change by the same term.
Therefore the solution 
    Eq.~(\ref{B.general.introduction})
contains
    {\it no absolute element}.~---
    {\it No boundary conditions} 
at spatial infinity 
are needed for regular solutions in
    Eq.~(\ref{B.general.introduction}).


\subsection{Exact dragging of inertial axes}


     The solution Eq.~(\ref{B.general.introduction})
can be rewritten to show exact dragging 
of inertial axes explicitly.
For reasons of 
     {\it symmetry} 
under rotations and space reflection
relative to a gyroscope at the origin, 
a general velocity field of matter for $r_{PQ}$ fixed 
can contribute to the gyroscope's precession 
only through its term with  
$( \ell = 1, \, P = + )$ 
in the spherical harmonic decomposition.
This term is a toroidal vorticity field,
and it is equivalent to a rigid rotation  
with the angular velocity 
$\vec{\Omega}_{\, \rm matter}^{\, \rm equiv} (r),$ 
      see Sec.~\ref{symmetries.Mach}.
From 
  Eqs.~(\ref{def.B}, \ref{B.general.introduction})
we obtain
\begin{eqnarray}
\vec{\Omega}_{\rm gyro} 
&=& <\vec{\Omega}_{\, \rm matter}^{\, \rm equiv}>  \, \, \, 
\equiv \int_0^\infty dr \, \,
\vec{\Omega}_{\, \rm matter}^{\, \rm equiv} (r) \, \, W(r),
\label{Mach.average.first.eq} 
\\
W(r) 
&=& \frac{1}{3} \, 16 \pi \, G_{\rm N} (\rho +p) \, 
R^3 \, Y_{\mu}(r)
\label{Mach.average.introduction}
\end{eqnarray}
for perturbations of 
open FRW universes. 
$Y_{\mu}(r)$ is the Yukawa force given in
    Eq.~(\ref{Yukawa.force.introduction}).


The crucial equation 
for establishing 
   {\it exact} 
(as opposed to partial) dragging
of local inertial axis directions is
the condition that the 
   {\it weight function} 
$W(r)$ 
must be 
     {\it normalized} 
to 
     {\it unity,}
\begin{equation}
\int_0^\infty dr \, W(r) = 1,
\label{weight.normalization.introduction}
\end{equation}
as it must be for a 
     {\it proper averaging weight function} 
in 
     {\it any} 
problem.
Our weight function with its Yukawa force,
     Eq.~(\ref{Mach.average.introduction}), 
fulfills
     Eq.~(\ref{weight.normalization.introduction})
as shown in
     Sec.~\ref{Mach.open.universe}.


     Eqs.~(\ref{Mach.average.first.eq},
           \ref{weight.normalization.introduction})
state our most important result:
$\vec{\Omega}_{\rm gyro}$ 
is equal to the weighted average 
(with proper normalization to unity)  
of $\vec{\Omega}_{\rm matter} (r)$
for all types of matter-energy and 
for all types of energy-current distributions.
The time-evolution of inertial axes 
     {\it exactly follows}
the weighted average
of cosmic matter motion.
Hence the evolution of inertial axes is 
    {\it fully determined}
(not merely influenced)
by the cosmic energy currents.
There is 
     {\it exact dragging} 
(not merely partial dragging)   
of inertial axes
by the 
     {\it weighted average} 
of cosmic energy currents, 
as postulated by Ernst Mach 
      \cite{Mach.Mechanik, Mach.Energy}.
Mach had asked:
\begin{quote} 
 ``What share has every mass in the determination of direction ... 
   in the law of inertia?
   No definite answer can be given by our experiences.''
\end{quote}
%


Exact dragging 
of inertial axes by arbitrary cosmological
energy currents in linear perturbation theory
has been demonstrated for the first time in
our papers
    \cite{Rovaniemi, CS.I}
for $K=0$
and in the present paper
for $K = \pm 1.$


\subsection{Alternative, inequivalent postulates proposed 
by authors after Mach}
\label{inequivalent.postulates}


Our presentation of Mach's principle
closely agrees with the one of 
   Misner, Thorne, and Wheeler
      \cite{MTW.Mach}
   and Weinberg
      \cite{Weinberg}.
However, 
    many alternative and inequivalent postulates 
have been proposed under the heading 
``Mach's principle'' by many authors
after E. Mach.
    An incomplete list 
of 10 inequivalent versions of 
Mach's principle  
has been given by Bondi and Samuel
   \cite{Bondi.Samuel},
see also  
   \cite{Barbour.Pfister}.


In the list of Bondi and Samuel 
   the version 
     ``Mach~9'' (``The theory contains no absolute elements'')
   is the postulate of the so-called ``relativists'': 
   Huyghens, Bishop Berkeley, Leibniz, Mach, and others.
   It was the starting point for Mach. 
   See our ``Test 3'' in 
       Sec.~\ref{fundamental.tests}.~---
The version 
      ``Mach~10'' (``Overall rigid rotations of a system are unobservable'')
is a necessary condition, 
formulated by Mach in his earliest writings
     \cite{Mach.Energy}
quoted in
   Sec.~\ref{Conclusions}.
See our ``Test 4'' in
     Sec.~\ref{fundamental.tests}.    
Globally rigid rotations are discussed in 
       Sec.~\ref{Globally Rigid Rotations: Zero Modes}.


In contrast the versions 
   ``Mach~1,~2,~6,~7''   
are not true 
  in cosmological General Relativity,
  as already stated in
   \cite{Bondi.Samuel}.~---
The version
   ``Mach~3'' (``local inertial frames influenced ...'') 
is irrelevant,
because with only partial dragging
     one violates
     Mach's postulates ``no absolute element'' in the input
     and ``frame-invariance'' in the solutions
     (see the quote of Mach in Sec. I.A).~---
   ``Mach~4'' (``The Universe is spatially closed'') 
is irrelevant, 
  because we show that 
  exact dragging also holds in open universes.~---
   ``Mach~5'' (``The total angular momentum of the universe is zero'') 
is not true 
  for that angular momentum which can be defined 
  and is observable  
  without an absolute element in the input
  (i.e. without prior knowledge
         of the local non-rotating frames), 
  see 
       Sec.~\ref{exponential.cutoff} 
  and the end of 
       Sec.~\ref{momentum.constraint}.~---
   ``Mach~8'' (``$\rho/\rho_{\rm crit}$ is of order unity'') 
is irrelevant, 
   because we show in 
       Secs.~\ref{Green.closed.universe}, \ref{Milne.deSitter}
that exact dragging also holds for $\rho/\rho_{\rm crit} \gg 1$
and $\rho/\rho_{\rm crit} \ll 1.$
    

With respect to this list of 10 versions of Mach's principle
we conclude:
The criticisms  
``There are many formulations of Mach's principle''
and 
``It is not clear, what Mach meant'' do not apply.


\subsection{Outline}


In 
   Sec.~\ref{momentum.constraint}
we derive the momentum constraint for 
(1) that momentum density which can be measured as an input 
without prior knowledge of
the solution of the momentum constraint (the output),
and for 
(2) the Laplacian $\Delta$ of de Rham-Hodge as opposed to the 
``rough Laplacian'' $\nabla^2.$


In 
   Sec.~\ref{Riemann.Efield.Bfield}
we show that
our $\vec{B}_{\rm g}$ field on a Minkowski background
uniquely determines 
all nonzero components of the Riemann tensor.


To derive the 
    {\it Green function} 
for the momentum constraint
for the curved 3-spaces $S^3$ and $H^3,$
we need the solution
in source-free regions, 
i.e. we need vorticity eigen-fields of the Laplacian.


In a first step, in 
     Sec.~\ref{eigenfcts.Laplacian.scalar.fields}, 
we derive 
   {\it scalar}
eigenfunctions of the Laplacian on $S^3$ and $H^3.$ 
Our new result: 
      We give the radial functions of the scalar harmonics in a 
             {\it simple, identical form}
      for 
             {\it all three geometries}
$(E^3, S^3, H^3).$


In 
     Secs.~\ref{vector.spherical.harmonics} - \ref{radial.fcts.vorticity}
we give the expansion of 
vorticity fields (toroidal and poloidal)
in terms of 
     {\it vector spherical harmonics} 
$\vec{X}^{\pm}_{\ell m}$
and $Y_{\ell m} \vec{e}_{\chi}.$
We show that the radial eigenfunctions 
of the Laplacian for 
     {\it toroidal} 
vector fields are the 
spherical Bessel, Neumann, and Hankel functions
generalized to $S^3$ and $H^3$ by the simple replacements 
$r \Rightarrow (\chi \, \, {\rm resp} \, \,  R)$ given 
     in Sec.~\ref{introduction.solution}.


In 
     Sec.~\ref{Sec.Mach}  
we show that only toroidal 
$\vec{A}_{\rm g}$-fields with $\ell = 1$ 
are relevant for the precession of gyroscopes at the origin and that
the precession of a gyroscope cannot be caused 
by scalar or tensor perturbations.
We then derive the radial Green functions 
for the operator $(\Delta - \mu^2)$
acting on toroidal  
$\vec{A}_{\rm g}$-fields with $\ell =1$
on $(E^3, S^3, H^3).$
The final result is the gravitomagnetic  
$\vec{B}_{\rm g}$-field at the position of the gyroscope,
i.e. the precession of the gyroscope, expressed by the 
energy currents in the universe, 
and the equation which expresses 
exact dragging of inertial axes.
We also show that Mach's principle remains valid
for perturbations of a universe
with arbitrarily small total energy
inside the Hubble radius,
i.e. arbitrarily close to the Milne universe,
and for a universe arbitrarily close to a de Sitter universe.~---
We derive the gravitomagnetic fields 
for the zero modes of the operator $(\Delta + 4K),$
i.e. for time-dependent globally rigid rotations 
of coordinates and FIDO axes in a FRW universe.~---
In 
      Sec.~\ref{fundamental.tests}
we give 
   {\it Six Fundamental Tests for Mach's Principle,}
all intimately related to Mach's original formulation 
and all explicitely fulfilled by our solutions.


The
      Appendix~\ref{appendix.diff.forms} 
gives the tools for  the de Rham - Hodge Laplacian 
on vector fields in Riemannian 3-spaces, 
specifically the explicit equations connecting 
the calculus of differential forms
with the elementary notation with curl etc.
These tools are essential for cosmological gravitomagnetism.


\section{THE MOMENTUM CONSTRAINT}
\label{momentum.constraint}



The momentum constraint written in  
    {\it local orthonormal tetrad} components
is given in 
     Eq.~(\ref{Einstein.equation.A.introduction}).
It can be derived via two different general methods:
(i) One can use the local orthonormal tetrad 
    (LONB) method 
    of Cartan from beginning to end. 
    This method has been presented 
    in Secs.~V and VI of
       \cite{CS.I},
    where the connection coefficients for vorticity perturbations 
    are given 
    for FRW backgrounds with $K=(0, \, \pm 1),$
    and $G_{\hat{0} \hat{i}}$
    is derived for $K=0.$
    We have used this method to derive 
       Eq.~(\ref{Einstein.equation.A.introduction})
    also for $K = \pm 1,$
    but we shall not present details again.
(ii) Alternatively one can, in a first step, 
    take the detour via the standard
    coordinate-component method with Christoffel symbols
    to obtain the momentum constraint in a 
       {\it coordinate basis} 
    with $T^0_{\, \, i}$ as the source.  
    $R^0_{\, \, i}$ has been derived in
       Bardeen~\cite{Bardeen},
    Eq.~(A2b), and in
       Kodama and Sazaki~\cite{Kodama.Sazaki},
    Eq.~(D.14b). 
    In a second step one applies the basis transformation 
    from the coordinate basis
    to the local orthonormal basis.~---  
A modification of method (ii) is quicker: 
The basis transformation at the end of method (ii) 
can be replaced by
taking Bardeen's momentum constraint
and moving the difference  $(T^0_{\, \, i} - T^{\hat{0}}_{\, \, \hat{i}}),$
which is meaningful at the level of linear perturbation theory,
to the left-hand side of the momentum constraint.


The momentum constraint for $R^0_{\, \, i}$ 
of Bardeen and of Kodama and Sazaki
reads in our notation
\begin{equation}
(\nabla^2 + 2 K/a_{\rm c}^2) A^i_{\rm g} 
= 16 \pi G_{\rm N} (\rho_0 + p_0) v_{\rm c}^i,
\label{Bardeen.momentum.constraint} 
\end{equation}
where $a_{\rm c}$ denotes 
the curvature scale,
and $\nabla^2$ refers to the physical scale. 
The 3-vector index $i$ in $A^i_{\rm g}$ and $v_{\rm c}^i$
is lowered and raised by the 3-metric.~--- 
The zero modes of the operator $(\Delta + 4K/a^2_{c})$
correspond to time-dependent rigid rotations of the coordinates
as demonstrated in
    Sec.~\ref{Globally Rigid Rotations: Zero Modes}.
For non-zero modes of $(\Delta + 4K/a_{c}),$
our gravitomagnetic vector potential $A^i_{\rm g},$
which is equal to our shift vector field $\beta^i,$
is directly proportional to 
Bardeen's gauge-invariant amplitude $\Psi$
(a combination of shift amplitude and shear amplitude),
\begin{equation}
A^i_{\rm g} (x,t) = \beta^i (x,t) = - \Psi (t) Q^{(1)i} (x),
\label{our.shift.Psi}
\end{equation}
for a given wave number and polarization.
For zero modes there is no shear, and 
a nonvanishing gauge-invariant amplitude does not exist.~---  
Similarly for Bardeen's gauge-invariant velocity amplitude $v_{\rm c}$
\begin{equation}
v_{\rm c}^i  (x,t)=  v_{\rm c} (t)  Q^{(1)i} (x).
\end{equation}
%


To convert Bardeen's momentum constraint 
   Eq.~(\ref{Bardeen.momentum.constraint})
to our momentum constraint,
the 
   {\it first step} 
is replacing the ``rough Laplacian'' 
$\nabla^2 = \nabla^i \nabla_i$
by ``the'' Laplacian $\Delta$ (de Rham-Hodge Laplacian)
via the 
    {\it Weitzenb\"ock} formula. 
With the Ricci tensor
of the 3-space for FRW the Weitzenb\"ock formula gives
    Eq.~(\ref{Weitzenboeck.FRW}),  
\begin{equation} 
(\Delta - \nabla^{2})\vec{A} 
= - (2K/a_{c}^{2})   \vec{A}.
\end{equation}
%
Hence Bardeen's momentum constraint written in terms of the 
de Rham-Hodge Laplacian 
is 
\begin{equation}
(\Delta + 4 K/a_{\rm c}^2)  A^i_{\rm g}
= 16 \pi G_{\rm N} (\rho_0 + p_0) v_{\rm c}^i.
\label{Bardeen.constraint.Laplacian}
 \end{equation}
%


The 
    {\it second step}   
is replacing 
the gauge-invariant matter velocity amplitude $v_{\rm c},$ 
used by Bardeen, 
by his other gauge-invariant amplitude $v_{\rm s},$
used by us. In our gauge, which is shear-free, $dx^i/dt$  
is equal to Bardeen's gauge-invariant $v^i_{\rm s}.$


The replacement of  $v_{\rm c}$ by  $v_{\rm s}$  
is necessary, because 
the field with the amplitude $v_{\rm c},$
used by Bardeen,
is directly related to the angular velocity of matter
    {\it relative to spin axes of local gyroscopes} 
at the sources throughout the universe.
This is 
     {\it not} 
a measurable input without 
  {\it prior knowledge of the solution of Einstein's equations}, 
i.e. of $g_{0i}, \, \vec{A}_{\rm g},$ and $\vec{B}_{\rm g}.$~---
In contrast, the field with the amplitude $v_{\rm s},$
used by us,
is directly related to the angular velocity of matter
measured relative to
geodesics (on $\Sigma_t$)
which start at our position and go in the directions of 
the chosen spatial axes along our world line
(resp geodesics to
asymptotic quasars 
in an open asymptotically unperturbed FRW universe).
The input needed 
in the context of Mach's principle
is the field with the amplitude $v_{\rm s}$.~---
The difference between 
Bardeen's two gauge-invariant velocity amplitudes is 
\begin{equation}
(v_{\rm c} -  v_{\rm s})  = - \Psi.
\label{vc.minus.vs}
\end{equation}
The $\Psi$ term 
is a geometric output, hence it must be moved 
to the left-hand side, the geometric side (output side), 
of the momentum constraint.
For the prefactors on Bardeen's right-hand side 
we use the ``second'' Friedmann equation,
\begin{equation}
(dH/dt) - \frac{K}{a_c^2} 
= - 4 \pi G_N (\rho +p).
\end{equation}
%
Combining the first and second steps gives
the new operator on the left-hand side (geometric output side)
of the momentum constraint: 
$[\Delta + (2+2-4)K/a_{\rm c}^2 - \mu^2] 
=[\Delta - \mu^2],$
i.e. the three 
     {\it curvature terms cancel}.
The new term on the right-hand side (measured input side) is 
our energy current $\vec{J}_{\epsilon}.$
This gives 
    Eq.~(\ref{Einstein.equation.A.introduction}).


From
   Eqs.~(\ref{vc.minus.vs},
         \ref{our.shift.Psi})
we obtain
$v_{\rm s}^i = dx^i/dt = v_{\rm c}^i - A_{\rm g}^i.$
Therefore the difference between the momentum densities
$(\rho + p) v_{\rm c}^i,$ used by Bardeen, 
and $(\rho + p) v_{\rm s}^i,$ used by us,
is analogous to the important difference
between the 
   {\it canonical} 
momentum $p_i = \partial L / \partial \dot{x}^i$
and the 
   {\it kinetic} 
momentum $k^i = m \, d x^i/dt$
in Lagrangian mechanics for point particles of charge $q$ 
in an electromagnetic field,
\begin{eqnarray}
k^i         
&=& p^i - q A^i,
\\
(\rho + p) \, v_{\rm s}^i 
&=& (\rho + p) \, [v_{\rm c}^i - A_{\rm g}^i],
\label{kinetic.canonical}
\end{eqnarray}
where the 3-indices are raised and lowered by the 3-metric.
The 
    {\it kinetic} 
(resp. 
    {\it canonical}) 
momentum 
    {\it is} 
(resp. 
    {\it is not}) 
measurable without prior knowledge of $A^i$ and $A^i_{\rm g}.$
The kinetic momentum density is equal to the LONB components
$T^{\hat{0}}_{\, \, \hat{i}} = (\rho + p) dx_{\hat{i}}/dt, $
while the canonical momentum is equal to the coordinate-basis components
$T^0_{\, \, i} = (\rho + p) \,
g^{(3)}_{ij} \, [dx^j/dt + A^j_{\rm g}].$


To prove the postulate formulated by Mach 
(exact dragging of inertial axis directions)
the kinetic momentum, resp the kinetic angular momentum of matter
is the appropriate input.
The canonical angular momentum (e.g. around the $z$-axis: $T^0_{\, \, \phi}$),
cannot be used, 
because its measurement resp. definition needs 
an absolute element in the input.~---
More details on kinetic versus canonical momentum and angular momentum 
in Sec.~VIII of
    \cite{CS.I}.


A cosmological term in Einstein's equations is diagonal in a LONB,
hence a cosmological constant cannot appear 
in the momentum constraint in LONB components.
If a cosmological constant $\Lambda$ 
is absorbed in $\rho$ and $p,$ 
one has $(p_{\Lambda} / \rho_{\Lambda}) = -1,$
and evidently $\Lambda$ cannot make a contribution to 
$\vec{J}_{\varepsilon} = (\rho + p) \vec{v}.$


\section{Riemann TENSOR UNIQUELY GIVEN BY OUR 
    $\vec{B}_{\rm g}$ FIELD}
\label{Riemann.Efield.Bfield}


In this section 
we show that our choice of FIDOs
and hence our $(\vec{E}_{\rm g}, \vec{B}_{\rm g})$ fields
are singled out uniquely.
We specialize to gravitomagnetism on a Minkowski background,
since the material in this section will not be needed later, 
although it is very important:
    {\it Our specific choice of FIDOs} 
is 
    {\it singled out}
by the property that  
    {\it only} 
with our choice of FIDOs   
the $(\vec{E}_{\rm g},  \vec{B}_{\rm g}  )$-fields   
    {\it uniquely determine} 
the 
    {\it Riemann tensor}   
and the gravito-electric and gravito-magnetic 
    {\it tidal tensors}  
$({\cal E}_{\hat{i} \hat{j}}, {\cal B}_{\hat{i} \hat{j}})$         
for vector perturbations,
\begin{eqnarray} 
- \,  R_{\hat{i} \hat{0} \hat{j} \hat{0}} 
&=& {\cal E}_{\hat{i} \hat{j}}
= E^{\rm g}_{(\hat{i} | \hat{j})},
\label{Riemann.dE}            
\\       
   -2 (^{\star}R)_{\hat{i} \hat{0} \hat{j} \hat{0}}    
   \equiv \varepsilon_{\hat{i} \hat{p} \hat{q}}
   \, R_{\hat{p} \hat{q} \hat{j} \hat{0}} 
   &=&   
   { \cal B}_{\hat{i} \hat{j}}
 =  B^{\rm g}_{(\hat{i} | \hat{j})},
 \label{Riemann.dB} 
\\
R_{\hat{k} \hat{0}} 
\equiv \delta_{\hat{r}\hat{s}} R_{\hat{k}\hat{r}\hat{0}\hat{s}} 
 &=& \frac{1}{2} (\mbox{curl} \vec{B}_{\rm g})_{\hat{k}}.
\label{Ricci.dB}
\end{eqnarray}
%
The purely spatial Riemann tensor is zero for vector perturbations,
i.e. the intrinsic geometry of each slice $\Sigma_t$ 
remains unperturbed,
as explained at the beginning of 
    Sec.~\ref{Vorticity.Perturbations}.~---  
A vertical bar  denotes the covariant derivative in 3-space, 
and  a round bracket around two indices denotes symmetrization.~---
The equation for $R_{\hat{k} \hat{0}}$ is 
Eq.~(35) of our 
     companion paper~\cite{CS.I}.


The 
     {\it first} 
     equalities in 
     Eqs.~(\ref{Riemann.dE}, \ref{Riemann.dB}) 
directly follow from the operational definition of the Riemann tensor:
(1) 
$R_{\hat{i} \hat{0} \hat{j} \hat{0}}$ 
gives  the
        {\it relative acceleration} 
of neighboring free-falling, quasistatic particles 
(geodesic deviation), 
i.e. the  
       {\it gravito-electric tidal field}
${\cal{E}}_{\hat{i}\hat{j}}$.
The tidal 3-tensors 
$({\cal{E}}_{\hat{i} \hat{j}}, {\cal{B}}_{\hat{i} \hat{j}})$
must be traceless by definition.
The 3-trace of $R_{\hat{i}\hat{0}\hat{j}\hat{0}}$ is a 3-scalar, 
therefore it is zero in the vector sector.~---
(2) The Hodge-star dual 
$(-^{\star}R)_{\hat{i} \hat{0} \hat{j} \hat{0}}
\equiv - \frac{1}{2} \varepsilon_{\hat{i} \hat{0} \hat{p} \hat{q}}
R_{\hat{p} \hat{q} \hat{j} \hat{0}}
  \equiv \frac{1}{2} \varepsilon_{\hat{i}         \hat{p} \hat{q}} 
R_{\hat{p} \hat{q} \hat{j} \hat{0}}$ 
gives the 
        {\it relative precession} 
of neighboring quasistatic gyroscopes' spins.
The 3-trace of 
$(\varepsilon_{\hat{i} \hat{p} \hat{q}} R_{\hat{p} \hat{q} \hat{j} \hat{0}})$
is zero because of the cyclic identity of the Riemann tensor,  
$R_{abcd} +R_{acdb} +R_{adbc} \equiv 0.$ 
Replacing the relative precession  
by the 
     {\it gravito-magnetic tidal field}
${\cal{B}}_{\hat{i}\hat{j}}$
gives a factor $(- \frac{1}{2})$
according to 
    Eq.~(\ref{def.B}) for any choice of FIDOs.


The 
    {\it second} 
equalities in 
     Eqs.~(\ref{Riemann.dE}-\ref{Ricci.dB})   
are only true for our $(\vec{E}_{\rm g}, \vec{B}_{\rm g})$-fields, 
defined via 
    {\it our specific choice of FIDOs.}
Our FIDOs are  
    {\it singled out} 
by the property that all 13 nonzero Riemann components  
are 
     {\it uniquely determined} 
by spatial derivatives of our 2 vector fields
$(\vec{E}_{\rm g}, \vec{B}_{\rm g}).$ 
If one also uses time-derivatives of the $\vec{B}_{\rm g}$ field
and uses ${\mbox{curl}} \vec{E}_{\rm g} = -\partial_t \vec{B}_{\rm g}$ 
from 
   Eq.~(29) in
   \cite{CS.I},
one concludes that
    {\it  our} 
$\vec{B}_{\rm g}$ field by itself 
    {\it determines} 
all 13 nonzero components of the 
    {\it Riemann} tensor:
Our $\vec{B}_{\rm g}$ field 
gives an efficient and physically transparent way 
to represent the entire information in the Riemann tensor.


The 
     {\it Weyl tensor} 
$C_{\alpha \beta \gamma \delta}$
(defined by having no nontrivial 4-contractions)
and its Hodge dual $^{\star} C,$
have been used in the 
     literature~\cite{Thorne.tidal.tensor, Bruni, Bertschinger}
to define the electric and magnetic tidal 3-tensors
by contracting twice with a time-like unit vector.
However, our tidal tensors have a 
     {\it different normalization,} 
because tidal tensors must be 
directly connected 
to relative acceleration 
and relative precession 
(and hence to relative gravito-magnetic fields). 
The normalization of our tidal tensors 
in terms of the Weyl tensor is given by 
$C_{\hat{i}\hat{0}\hat{j}\hat{0}} = \frac{1}{2} {\cal{E}}_{\hat{i}\hat{j}}$
and 
$^{\star}C_{\hat{i}\hat{0}\hat{j}\hat{0}} = 
-\frac{1}{2} {\cal{B}}_{\hat{i}\hat{j}}.$
The factor $\frac{1}{2}$ in the first equation comes from 
$C_{\hat{i}\hat{0}\hat{j}\hat{0}} = 
\frac{1}{2} R_{\hat{i}\hat{0}\hat{j}\hat{0}}$ 
valid for 
     {\it vector} 
perturbations. 
The factor $(-\frac{1}{2})$ in the second equation 
comes from the conversion
from the relative precession 
(given by the Riemann tensor) to the 
relative gravito-magnetic field.


The 
    {\it curvature invariant} 
is  
${\bf R \cdot R} \equiv 
R_{\alpha \beta \gamma \delta}  R^{\alpha \beta \gamma \delta} = 
  8 \, [ \, {\cal E}_{\hat{i} \hat{j}}  {\cal E}_{\hat{i} \hat{j}} 
- \frac{1}{4} {\cal B}_{\hat{i} \hat{j}}  {\cal B}_{\hat{i} \hat{j}} \, ],$
and  the curvature pseudoinvariant is ${\bf ^{\star} R  \cdot R} 
\equiv (\frac{1}{2} \varepsilon_{\alpha \beta \rho \sigma} 
R^{\rho \sigma}_{\, \, \, \, \, \gamma \delta})  
R^{\alpha \beta \gamma \delta} 
=  - \, 4 \, {\cal E}_{\hat{i} \hat{j}}  {\cal B}_{\hat{i} \hat{j}}.$
These invariants are 
     {\it not useful} 
for first-order perturbation theory,
because local Lorentz transformations of 
${\cal{E}}_{\hat{i}\hat{j}}$ and 
${\cal{B}}_{\hat{i}\hat{j}}$ 
are 
    {\it second-order} 
in perturbations.


The connection in the opposite direction, finding our 
$\vec{B}_{\rm g}$ field from the Ricci tensor: 
Any vector field in the 3-vector sector, i.e. with zero divergence, 
is the sum of 
a field determined by its curl plus a harmonic field,
which in Euclidean 3-space is a homogeneous field, 
because we require that the $\vec{B}_{\rm g}$-fields 
are not singular at spatial infinity. 
Therefore the Ricci tensor determines our  $\vec{B}_{\rm g}$-field
up to a homogeneous $\vec{B}_{\rm g}$-field, 
which is equivalent to a 
time-dependent spatially rigid rotation of all FIDOs according to 
     Eq.~(\ref{def.B}).~--- 
Our solutions of the equations of cosmological gravitomagnetism
on FRW backgrounds with $K=(0, \pm 1),$ e.g.
     Eq.~(\ref{B.general.introduction}),
are   
    {\it form-invariant} 
under time-dependent spatially rigid rotations,
    see Sec.~\ref{sect.frame.invariance}.
The exact dragging of inertial axes,
      Eq.~(\ref{Mach.average.first.eq}),
is 
     {\it independent of the choice of FIDOs}.


Before Einstein's General Relativity, 
gravitomagnetic and gravitoelectric fields 
were discussed by Oliver Heaviside
    \cite{Heaviside}
based on a gravitational-electromagnetic analogy.
With his equations Heaviside could have derived 
the partial dragging of inertial axes
by a rotating star (Lense-Thirring effect), 
although a factor 4 would have been missing.


\section{EIGENFUNCTIONS OF THE LAPLACIAN FOR SCALAR FIELDS AND 
VORTICITY FIELDS IN $(E^3, S^3, H^{3})$}
\label{eigenfcts.Laplacian.vorticity.fields}


Cosmological vorticity fields  
are of a different type 
from vorticity fields typical in fluid dynamics 
and condensed matter physics
with their vortex lines (line singularities) 
and fields often idealized 
with cylindrical symmetry.
In cosmology we consider vorticity fields 
($\equiv$ vector fields with vanishing divergence) 
which are 
    {\it regular,}
and we expand them in a 
    {\it spherical basis.} 
These same fields also occur in the
multipole expansion of electromagnetic radiation fields
      \cite{Jackson.vector.spherical.harmonics}.
In this section we generalize these fields to the
    {\it curved 3-spaces} 
$S^3$ and $H^3.$~---
In the first paragraph of 
       Sec.~{\ref{vector.spherical.harmonics}
we explain, why the methods used in textbooks 
and in the literature for CMB-analyses are not adapted for our purpose.       


Our aim is to solve the momentum constraint
with its operator $(\Delta - \mu^2),$ 
i.e. the inhomogeneous Helmholtz equation.
As a preparation for the Green function of
$(\Delta - \mu^2)$ 
we derive the eigenfunctions of the Laplacian
for $(E^3, S^3, H^3)$
inside resp. outside a shell with sources.
First we do this for scalar fields.


\subsection{Eigenfunctions of the Laplacian for scalar fields 
in {\bf $(E^3, S^3, H^3)$}}
\label{eigenfcts.Laplacian.scalar.fields}


Our new results in this subsection:
      We give the radial functions of the scalar harmonics in a 
             {\it simple, identical form}
      for 
             {\it all three geometries}
$(E^3, S^3, H^3),$ e.g.
             Eq.~(\ref{radial.solution.1}),           
a form which is familiar 
from the spherical Bessel functions for elementary wave physics 
in Euclidean 3-space.
This is in contrast to the complicated derivations 
and results in the literature.~---
      In Subsec.~\ref{radial.fcts.vorticity}
      we shall show that the radial functions are also identical for 
      scalar and 
             {\it toroidal vorticity} 
      harmonics, if and only if one uses 
      those vector spherical harmonics,  
      $\vec{X}^{-}_{\ell m},$
      which have point-wise norm and  LONB components 
      independent of $R$ for fixed $(\theta, \phi).$


Since the momentum constraint
    Eq.~(\ref{Einstein.equation.A.introduction})
is an elliptic equation (no partial time-derivatives),
we consider,
in this and all following sections, 
a FRW universe with $K=\pm 1$ at a
    {\it given time} 
with 
    {\it units of length} 
chosen such that the curvature radius is
$a_{c}=a=1$ at the given time.
We also include the case of 
FRW with $K=0$ (Euclidean 3-space $E^3$) 
at a given time with the scale factor $a=1,$
\begin{eqnarray}
a &=& 1
\\
ds^2 &=& d\chi^2 + R_K^2(\chi) \, [d\theta^2 + 
\sin^2 \theta \, d \phi^2]
\label{line.element}
\\
R_K(\chi) &=& \{\chi, \, \sin \chi, \, \sinh \chi  \}, 
\quad 
K = \{0, \, +1, \, -1 \}. \, \, \, \, \, \, 
\end{eqnarray}
The following relations will be useful,
\begin{equation}
R^{\prime \prime}= - K R, \quad \quad R^{\prime \, 2} = 1 - K R^2,
\label{R.double.prime}
\end{equation}
where $^{\prime} = d/d \chi,$   
and  $R$ is short for $R_K(\chi).$


For scalar fields 
the angular eigenfunctions of the Laplacian 
$\Delta$
are spherical harmonics
$Y_{\ell m}(\theta, \phi)$ with eigenvalues
$ - \ell(\ell + 1)/ R^{2}.$~--- 
The radial eigenfunctions of the Laplacian
are the generalization to $K = \pm 1$ of the
spherical Bessel functions $j_{\ell}(qr),$
which are the radial eigenfunctions 
in Euclidean 3-space.
These 
    {\it generalized spherical Bessel functions}, 
the hyperspherical Bessel functions,
will be denoted in this paper
by $\tilde{\jmath}^{\, (K)}_{q \ell} (\chi)$ 
to emphasize the close correspondence to
$\tilde{\jmath}^{\, (K=0)}_{q \ell} (\chi)
\equiv j_{\ell}(q \chi).$ 
We shall always denote the
    {\it eigenvalues} 
of the de Rham-Hodge Laplacian $\Delta$
by $(-k^2).$ 
For $\Delta$ acting on scalar eigenfunctions
and on vector or tensor eigenfunctions in the scalar sector 
we define $q$ by
\begin{equation}
\mbox{scalar sector:}  \quad \quad q^2 = k^2 + K.
\label{eigenvalues}
\end{equation}
It will turn out that $q$ is the 
    {\it wave number} 
in the 
    {\it radial oscillations}
$\sin (q\chi)$ in
   Eqs.~(\ref{radial.solution.1},
         \ref{radial.function.s-wave}, 
         \ref{radial.function.p-wave}).


The radial part of the Laplace operator 
acting on a scalar function $f(\chi)$ is
$ \Delta f = 
R^{-2} \, \frac{d}{d\chi} 
(R^{2}  \, \frac{d}{d\chi} f).$
Using $R^{-1} \frac{d^2}{d \chi ^2} R=-K$
we obtain $\Delta f = R^{-1} \frac{d^2}{d \chi ^2} (Rf) + Kf.  $
Hence the radial eigenfunctions 
for scalar harmonics satisfy
\begin{equation}
\frac{1}{R} \,  \frac{d^2}{d\chi^2} (R \, 
\tilde{\jmath}_{q \ell}^{(K)}) 
= [- q^{2} + \frac{\ell(\ell + 1)}{R^2}] \,
\tilde{\jmath}_{q \ell}^{(K)}.
\label{radial.equation}
\end{equation}
The solutions regular at the origin are 
\begin{equation}
\tilde{\jmath}_{q \ell}^{(K)} (\chi)=
R^{\ell} \, \,  
(- \frac{1}{R} \, \frac{d}{qd\chi})^{\ell} 
(\frac{\sin q\chi}{qR}),
\label{radial.solution.1}
\end{equation}
which we call 
         {\it generalized spherical Bessel functions.}
Writing the radial equation and the solutions in this form 
gives a 
         {\it unified notation} 
for all 
         {\it three geometries} 
($E^3, S^3, H^3).$ 
Compared to spherical Bessel functions 
(familiar from wave physics
in Euclidean 3-space)
the only changes are the replacements
of the 
           {\it independent variable,} 
$r \rightarrow \chi,$ 
in the radial oscillations and in the derivative,
and the replacement of the 
           {\it denominators and prefactors,} 
$r \rightarrow R(\chi).$~---
From 
     Eq.~(\ref{radial.solution.1})
we obtain the generalized spherical Bessel functions  
$\tilde{\jmath}_{q \ell}^{(K)} (\chi) $ 
for $\ell=(0, 1),$    
%
\begin{eqnarray}
\tilde{\jmath}_{q,\ell = 0}^{\, (K)}(\chi)
&=& \frac{1}{qR}   \sin \, (q\chi),
\label{radial.function.s-wave}
\\
\tilde{\jmath}_{q,\ell = 1}^{\, (K)}(\chi) 
&=& 
- \, \frac{1}{q^2} \, \frac{d}{d \chi} 
[\frac{1}{R} \sin (q \chi)].
\label{radial.function.p-wave}
\end{eqnarray}
%
The generalized spherical Neumann functions
are obtained from 
   Eqs.~(\ref{radial.solution.1}, 
         \ref{radial.function.s-wave},
         \ref{radial.function.p-wave})
by the replacements
$\sin (q \chi) \rightarrow -\cos (q \chi)$  and
$\cos (q \chi) \rightarrow +\sin (q \chi).$
The generalized spherical Hankel functions
are given by $\tilde{h}^{(1,2)} = \tilde{j} \pm i \tilde{n}.$~---
Relevant for Mach's principle are the 
spherical Bessel and Hankel functions 
for $\ell = 1$ generalized to $K = \pm 1.$
They are given by 
     Eq.~(\ref{radial.function.p-wave})
and
\begin{equation} 
\tilde{h}_{q,\ell = 1}^{(1,2)(K)}(\chi) 
= 
\pm i \, \frac{1}{q^2} \, \frac{d}{d \chi} 
[\frac{1}{R} \exp (\pm i q \chi)].
\label{Hankel.p-wave}
\end{equation}
%


The eigenfunctions of the Laplacian will be used
in the context of Green functions with sources at $\chi = \chi_{\rm s},$
hence they will be used only for 
$0 \le \chi  < \chi_{\rm s}$ resp. for
$\chi > \chi_{\rm s} .$ 
Therefore there are 
   {\it no geometric restrictions on the eigenvalues}
in our context.


The 
      {\it simplicity} 
of the elementary functions
     ~(\ref{radial.function.p-wave},
       \ref{Hankel.p-wave}),
their
      {\it identical form} 
for all three geometries, 
the 
      {\it familiar appearance} 
from elementary wave physics in Euclidean 3-space,
and the 
      {\it straightforward derivation}
are in contrast to the complicated derivations
for the hyperspherical Bessel functions $\Phi^{\nu}_{\ell} (\chi)$
via the associated Legendre functions 
$P^{-1/2-\ell}_{-1/2 + \beta}(\cos y)$ discussed extensively in 
    \cite{Abbott.Schaefer}
and used by                                  
    \cite{Hu.Seljak.White.Zaldarriaga.App.B}.


\subsection{Vector spherical harmonics}
\label{vector.spherical.harmonics}


In this subsection we introduce
the 3-dimensional vector spherical harmonics 
$(\vec{X}^{\pm}_{\ell m}, \vec{e}_{\chi} Y_{\ell m})$
and the Regge-Wheeler harmonics 
$(\tilde{x}^{\pm}_{\ell m}, \tilde{d}_\chi Y_{\ell m}),$ 
and we obtain their curls and divergences in 
     Eqs.~(\ref{div.x.plus}) - (\ref{curl.rad}),
which are valid and 
     {\it form-identical} 
not only for the 3-spaces $(S^3, H^3, E^3),$ 
but for 
    {\it any spherically symmetric Riemannian 3-space.}
Our curl- and 
    div-equations
for this 
    {\it general} 
case are 
    {\it very much simpler} 
than the equations in the literature 
for the special case of Euclidean 3-space.~---
The general and simple
    equations~(\ref{div.x.plus}) - (\ref{curl.rad})
are 
    {\it all that's needed}
to construct vorticity fields
(divergenceless) 
and to explicitly write down 
the de Rham-Hodge Laplacian on vorticity fields,
    $\Delta \vec{V} = -$ curl curl $\vec{V},$
for 
    {\it any} 
Riemannian 3-space with spherical symmetry.~---
Textbooks usually present the vector spherical harmonics of 
     \cite{Rose, Edmonds}, 
which are not adapted to cosmological perturbation theory,
because they mix the vector and the scalar sector, 
see the end of this subsection.~---
The spin-$s$ spherical harmonics used in CMB analyses 
      \cite{Hu.Seljak.White.Zaldarriaga}
are not adapted to our problem, 
because they mix parities, i.e. the toroidal and poloidal sectors 
of vector spherical harmonics,
see the end of this subsection and 
    Sec.~\ref{vorticity.fields}.


Vector spherical harmonics 
$\vec{X}^{\pm}_{\ell m}(\theta, \phi)$ 
form a basis for vector fields 
tangent to the 2-sphere.
In the notation of differential forms, 
the vector spherical harmonics of 
     {\it Regge and Wheeler}
     \cite{Regge.Wheeler}, 
which we denote by a 
     {\it lower case} 
$\tilde{x}^{\pm}_{\ell m},$ are given by
\begin{equation}
\tilde{x}_{\ell m}^{+}  \equiv  
d Y_{\ell m}, 
\quad 
\tilde{x}_{\ell m}^{-}  \equiv  \, 
- \,  ^{(2)}\ast \, \, d Y_{\ell m},   
\label{vector.spherical.harmonics.2dim}
\end{equation}
i.e. the 
      {\it gradient} 
of  $Y_{\ell m},$ and its 
      {\it Hodge dual} 
on the 2-sphere,  $^{(2)}\ast.$ 
In component notation 
     Eq.~(\ref{vector.spherical.harmonics.2dim}) 
gives  
\begin{eqnarray}
\quad  (x_{\ell m}^{+})_{\alpha} 
&=&  \partial_{\alpha} \,Y_{\ell m}, 
\label{vector.spherical.harmonics.2dim.even.parity}
\\
(x^{-}_{\ell m})_{\alpha} 
&=& - \varepsilon_{\alpha \beta} \, \, g^{\beta \gamma} \, 
\partial_{\gamma} \,Y_{\ell m},
\label{vector.spherical.harmonics.2dim.odd.parity.A}
\end{eqnarray}
where  $\alpha = (\theta, \phi), \,$
$\varepsilon_{\alpha \beta} = \sqrt{^{(2)}g} \, 
\varepsilon_{\hat{\alpha} \hat{\beta}}, \, $ 
$\varepsilon_{\hat{\theta} \hat{\phi}} \equiv +1.$ 
In the combination 
$\varepsilon_{\mu \nu} g^{\nu \lambda} = 
\varepsilon_{\mu}^{\, \, \, \, \lambda}$ 
the radius $R$ of the 2-sphere drops out,
$\varepsilon_{\theta}^{\, \, \, \, \phi} =  (\sin \theta)^{-1}, \,
 \varepsilon_{\phi}^{\, \, \, \, \theta} = -(\sin \theta). $


The vector harmonics of Regge and Wheeler 
have 
      {\it covariant components} 
(1-form components) which are  
independent of the radial coordinate 
$\chi,$
\begin{equation}
\partial_{\chi} (x^{\pm}_{\ell m})_{\alpha} =0,   
\quad  \quad \alpha = (\theta, \phi).
\end{equation}
%

In contrast, the 
      {\it 'physical'} 
vector spherical harmonics
are denoted by the 
      {\it upper case} 
$\vec{X}^{\pm}_{\ell m}.$
They are used in the literature on
classical electrodynamics 
   \cite{Jackson.vector.spherical.harmonics} 
and called pure-spin vector harmonics by Thorne
   \cite{Thorne.RMP}. 
They are defined with an $R$-factor 
which makes the 
     {\it radial covariant derivative} 
vanish,
\begin{eqnarray} 
\vec{X}^{\pm}_{\ell m}  &\equiv& R \, \, \vec{x}^{\pm}_{\ell m},
\label{physical.vector.spherical.harmonics}
\\
\nabla_{\chi} \, \vec{X}^{\pm}_{\ell m} &=& 0.
\label{pointwise.norm} 
\end{eqnarray}
This is equivalent to requiring that the 
     {\it point-wise norm}
$g(\vec{X}^{\pm \ast}_{\ell m}, \vec{X}^{\pm}_{\ell m})$
and the 
     {\it LONB components} 
of $\vec{X}^{\pm}_{\ell m}$
are independent of $R$ for fixed $(\theta, \phi).$


We rewrite the 2-dimensional Hodge dual of
   Eq.~(\ref{vector.spherical.harmonics.2dim}) as
$ ^{(2)}\ast \,  \vec{\nabla} Y_{\ell m} = 
- \vec{e}_{\chi} 
\times  \vec{\nabla} Y_{\ell m}.$
The generator of rotations, 
the angular momentum operator of wave mechanics 
in units of $\hbar,$ is
$\vec{L} = -i (\vec{R} \times \nabla),$ where 
$\vec{R} \equiv R(\chi)\vec{e}_{\chi}.$
Hence
\begin{eqnarray}
&\vec{X}_{\ell m}^{+} & = R  \vec{\nabla} Y_{\ell m}
\label{3d.spherical.harmonics.even}
\\
&\vec{X}_{\ell m}^{-} & = i \vec{L} Y_{\ell m}
= (\vec{R} \times         \vec{\nabla})Y_{\ell m}
=  \vec{e}_{\chi} \times \vec{X}^+_{\ell m}
\label{3d.spherical.harmonics.odd}
\\
&\vec{e}_{\chi} Y_{\ell m}, &
\end{eqnarray}
where we have also listed the third basis field
needed for 3 dimensions, $\vec{e}_{\chi} Y_{\ell m}.$
The triple $( \vec{X}^+, \, \vec{X}^-, \, \vec{e}_{\chi} )$
is point-wise orthogonal and has positive orientation.
$\vec{X}^{+}$ and $\vec{X}^{-}$ have the same norm point-wise.


The     {\it parity} 
is $P=(-1)^{\ell}$ 
for $\vec{X}_{\ell m}^{+}$ and $\vec{e}_{\chi} Y_{\ell m},$    
while 
$P=(-1)^{\ell +1}$
for $\vec{X}_{\ell m}^{-}.$


All three basis fields 
$\{ \vec{X}^{\pm}_{\ell m}, \vec{e}_{\chi} Y_{\ell m} \}$
are eigenfunctions of the total angular momentum operators
$\{ J^{2}, J_{z} \}$ with eigenvalues 
$\{ \ell(\ell +1), m  \}.$
This follows because
exterior differentiation $\tilde{d},$ taking the 
Hodge dual $(\ast),$ and multiplication with $\vec{e}_{\chi}$
commute with rotations of the total system.


The {\it normalization} and {\it orthogonality} relation 
on the 2-sphere at any $R$ is
\begin{eqnarray}
(\vec{X}_{\ell m}^{(p)}, \vec{X}_{\ell' m'}^{(p')}) 
&=& 
\int d\Omega <\vec{X}_{\ell m}^{(p)}, \vec{X}_{\ell' m'}^{(p')}> 
\nonumber 
\\
&=& \ell (\ell +1) \delta_{\ell \ell'} \delta_{m m'} 
\delta_{p p'}.
\label{norm}
\end{eqnarray}
This relation is crucial 
for the projection of a general velocity field on the 
term with $(\ell =1, P = +).$
The global (Hilbert space) scalar product on the 2-sphere 
is denoted by $(\vec{V},\vec{W}),$
and $<\vec{V},\vec{W}>$ denotes 
the point-wise scalar product     
  $g(\vec{V}^{\ast},\vec{W}),$ 
where the superscript $^{\ast}$ 
stands for the complex conjugate.


    {\it Computations}
of curl and div and 
    {\it results} 
are much 
    {\it simpler} 
for Regge-Wheeler harmonics 
$\vec{x}^{\, \pm}_{\ell m}$
than for the 'physical' harmonics $\vec{X}^{\pm}_{\ell m}.$---
Also the basis fields $[\vec{e}_{\chi} R^{-2} Y_{\ell m}]$ 
are computationally much simpler than 
the basis fields $[\vec{e}_{\chi} Y_{\ell m}],$
because the divergence of the former vanishes 
apart from a $\delta$-function at the origin, 
which turns out to be irrelevant in our context,
\begin{eqnarray}
\mbox{div} \, \vec{x}^{\, +}_{\ell m}
&=& - \, \frac{\ell (\ell +1)}{R^2} Y_{\ell m}
\label{div.x.plus}  
\\
\mbox{curl} \, \vec{x}^{\, +}_{\ell m}
&=& 0
\label{curl.x.plus}
\\
\mbox{div}  \, \vec{x}^{\, -}_{\ell m} 
&=& 0
\label{div.x.minus}
\\
\mbox{curl} \, \vec{x}^{\, -}_{\ell m} 
&=&  - \, \frac{\ell (\ell +1)}{R^2} Y_{\ell m} \vec{e}_{\chi}
\label{curl.x.minus}
\\
\mbox{div} \,   [Y_{\ell m} \frac{\vec{e}_{\chi}}{R^2}] 
& = & 0
\label{div.rad}
\\
\mbox{curl} \,  [Y_{\ell m} \frac{\vec{e}_{\chi}}{R^2}] 
&=& - \, \frac{1}{R^2} \vec{x}^{\, -}_{\ell m}.
\label{curl.rad}
\end{eqnarray}
If we had computed the divergence and curl of 
$\vec{X}^{\pm}_{\ell m}$ and $Y_{\ell m} \vec{e}_{\chi},$
there would have been twice as many terms, 
and the terms would have been more complicated.


Computations of  
     $\mbox{curl} \, \vec{V} = \vec{\nabla} \times \vec{V}$
and  $\mbox{div}  \, \vec{V} = \vec{\nabla}~\cdot~\vec{V},$ 
where $\vec{\nabla}$ is the covariant derivative,
are made very simple with 
the calculus of 
     {\it differential forms},
where no Christoffel symbols 
are needed, see 
     Eqs.~(\ref{curl},
           \ref{div}).


The 
     {\it spin-$s$ spherical harmonics} 
of Newman and Penrose
     \cite{Newman.Penrose.JMP, Goldberg, Thorne.RMP},
which are widely used for cosmic microwave anisotropies
     \cite{Hu.Seljak.White.Zaldarriaga},
are 
     {\it not adapted} 
to our problem, because they 
     {\it mix parities,}
i.e. they mix toroidal and poloidal vorticity fields
     (see Sec.~\ref{vorticity.fields}),
while the dragging of inertial axis-directions 
is caused by energy-flows with $J^P = 1^+,$ 
i.e. in the toroidal sector
     (see Sec.~\ref{symmetries.Mach}).


In most textbooks 
vector spherical harmonics are constructed by coupling a definite 
    {\it orbital angular momentum} 
with 
    {\it spin-1 basis vectors} 
to obtain a definite total angular momentum 
via Clebsch-Gordan coefficients
    \cite{Rose, Edmonds, Thorne.RMP}.
These basis states are 
    {\it not adapted} 
to cosmological perturbation theory,
because they 
   {\it mix} 
the 
   {\it poloidal vorticity sector} 
and the
   {\it scalar sector},
see the following subsection.


\subsection{Toroidal and poloidal vorticity fields}
\label{vorticity.fields}


For 3-dimensional vorticity fields the
3-divergence must be zero. 
The vector spherical harmonics 
$\vec{x}^{\, -}_{\ell m}$ for $m=0$
are vector fields purely along the direction of 
$\pm \, \vec{e}_{\phi}$   
and independent of $\phi.$  
Therefore the 
$\vec{x}^{\, -}_{\ell m}$ 
are named 
     {\it 'toroidal',} 
and evidently their divergence is automatically zero. 
The toroidal harmonics with $m \neq 0$ are generated
from those with $m = 0$ by rotations, 
and their divergence is again zero, 
as also seen from
   Eq.~(\ref{div.x.minus}).---
Multiplication with a radial function 
$g^{{\rm tor}} (\chi)$
gives the 3-dimensional 
toroidal vorticity fields, 
\begin{equation}
\vec{V}^{\, {\rm tor}}_{\ell m} (\chi, \theta, \phi)=  
g^{{\rm tor}}_{\ell} (\chi) \, \, 
\vec{x}_{\ell m}^{\, -} (\theta, \phi).  
\label{toroidal.vector.field}
\end{equation}
%

 
The 
    {\it 3-divergence}
is 
    {\it not zero} 
for the vector spherical harmonics   
$\vec{x}^{\, +}_{\ell m}$
by themselves, 
${\rm div} \, \vec{x}_{\ell m}^{\, +} \neq 0.$  
To obtain vector fields $\vec{V}$ 
with zero 3-divergence we must add 
a part perpendicular to $S^{2}, \quad$
$\vec{e}_{\chi} F(\chi, \theta, \phi),$
where $F$ can be expanded in 
scalar spherical harmonics $Y_{\ell m}.$
Hence the 3-dimensional 
      {\it poloidal vorticity fields} can be written
\begin{equation}
\vec{V}_{\ell m}^{{\rm pol}} (\chi, \theta, \phi) = 
  g_{\ell}^{({\rm pol, \, tg}) } \, \vec{x}_{\ell m}^{\, +}
+ g_{\ell}^{({\rm pol, \, rad})} \, [Y_{\ell m} R^{-2} \, \vec{e}_{\chi}].
\label{poloidal.vector.field} 
\end{equation}
Both terms have 
$P=(-1)^{\ell}.$
For $m=0$ this equation gives vorticity fields $\vec{V}$
with only radial and $\theta -$components, 
since $\vec{x}^{\, +}_{\ell, m=0}$ 
points along the meridians. 
Hence the name 'poloidal' for these vorticity fields.


The condition 
for vanishing divergence 
of the poloidal vorticity fields is evaluated using
the vector identity 
      $\mbox{div} \, (g \vec{V}) 
= g \, \mbox{div} \,    \vec{V} + \vec{V} \cdot \mbox{grad} \, g$
and
    Eqs.~(\ref{div.x.plus},
          \ref{div.rad}),
\begin{equation}
\frac{d}{d \chi} \, g_{\ell}^{({\rm pol, rad})}
=\ell (\ell +1)  \, g_{\ell}^{({\rm pol, tg})}.
\label{div.pol.zero}
\end{equation}
Again, the equations in this subsection are valid for 
   {\it any}
Riemannian 3-space with spherical symmetry.


\subsection{The radial functions for vorticity harmonics}
\label{radial.fcts.vorticity}


Relevant for rotational frame-dragging are 
energy currents and 
vector potentials in the 
     {\it toroidal} 
sector as demonstrated in
   Sec.~\ref{symmetries.Mach}. 
Hence we now focus on 
the toroidal eigenfunctions of the Laplacian.~---
     {\it Poloidal}
energy currents cannot contribute 
to the precession of a gyroscope at the origin.
Therefore poloidal eigenfunctions of the Laplacian 
are not considered in this paper, 
but they will be needed and presented
in a forthcoming paper, ``The other half of Mach's principle:
Frame dragging for linear acceleration''
    \cite{CS.III}.


In this subsection we derive the following  important 
and simple 
       {\it result:} 
\begin{enumerate}
  \item In the eigenfunctions of the Laplacian 
        for 
           {\it toroidal vorticity}  
        fields
        on $(S^3, H^3, E^3)$: 
        The 
           {\it physical} 
        vector spherical harmonics $\vec{X}^{-}_{\ell m}$ 
        are multiplied by 
           {\it radial functions} 
        which are 
           {\it identical} 
        to those for scalar harmonics,   
        i.e. the 
           {\it generalized spherical Bessel functions} 
        $\tilde{\jmath}_{q \ell}^{\, (K)} (\chi)$  
        of 
            Eq.~(\ref{radial.solution.1})
        resp. the corresponding Neumann and Hankel functions.
\item   For 
           {\it toroidal vorticity}  
        harmonics we have the relation $q = k$:
\begin{equation}
(\Delta + q^2) [\tilde{\jmath}_{q  \ell}^{(K)} (\chi) \, 
\vec{X}^-_{\ell m} (\theta, \phi)] = 0.
\label{main.result.radial.toroidal}
\end{equation}
        In all sectors 
        $q$ denotes the 
            {\it wave number in the radial oscillations} 
        $\sin (q \chi),$  
        and $(-k^2)$ denotes the 
            {\it eigenvalue of the de Rham-Hodge Laplacian.} 
        Note that the relation for scalar 
        harmonics is
        $q^2 = k^2 + K,$ 
            Eq.~(\ref{eigenvalues}).
\end{enumerate}

To prove this result, all that's needed is the expression for
the de Rham-Hodge Laplacian for vorticity fields, 
$\Delta = - \mbox{curl curl} \vec{A},$ 
and the simple Eqs.~(\ref{div.x.plus}) - (\ref{curl.rad}).~--- 
For the rest of this subsection we drop the subscripts $(\ell, m).$
We start with a toroidal vorticity field 
$\vec{A}^{\, {\rm tor}} = g_A^{{\rm tor}} \vec{x}^{\, -}$
and evaluate its curl. 
Using the vector identity 
$    \mbox{curl} \, (g \vec{v}) =
g \, \mbox{curl} \, \vec{v} + (\mbox{grad} \, g) \times \vec{v}$
and 
   Eq.~(\ref{curl.x.minus})
gives
\begin{eqnarray}
&& {\rm curl} \, \vec{A}^{\, {\rm tor}} 
= {\rm curl} \, (g_A^{{\rm tor}} \vec{x}^{\, -})
\nonumber
\\
&& \, \, \, \, \, \,   
= -  (\frac{d}{d\chi} g_A^{{\rm tor}}) \vec{x}^{\, +}    
- \ell(\ell +1)    g_A^{{\rm tor}} [Y R^{-2} \vec{e}_{\chi}].
\label{curl.A.toroidal}
\end{eqnarray}
%
In a second step we take the curl a second time, which gives 
minus the de Rham-Hodge Laplacian 
of $\vec{A}^{\, {\rm tor}},$
\begin{eqnarray}
\Delta \vec{A}^{\, {\rm tor}} 
&=& - {\rm curl \, curl} \, \vec{A}^{\, {\rm tor}}
\nonumber
\\
&=& 
[    \frac{d^2}{d\chi^2}     g_A^{{\rm tor}}
- \frac{\ell(\ell +1)}{R^2}   g_A^{{\rm tor}}] \vec{x}^{\, -},
\label{Laplacian.A.tor}
\end{eqnarray}
where we have used
    Eqs.~(\ref{curl.x.plus},
          \ref{curl.rad}).
Up to here all formulae in 
     Secs.~\ref{vector.spherical.harmonics} 
         - \ref{radial.fcts.vorticity}
are valid for 
     {\it any}
Riemannian 3-space with spherical symmetry.
We now specialize to $(S^3, H^3, E^3).$
We rewrite the toroidal vorticity fields 
in terms of the 
     {\it physical} 
vector harmonics
$\vec{X}^{-}_{\ell m} = R \, \vec{x}^{\, -}_{\ell m}$
and their associated radial functions  
$G^{{\rm tor}} \equiv R^{-1} g^{{\rm tor}},$
hence $g^{\rm tor} \vec{x}^- = G^{\rm tor} \vec{X}^-.$
Comparing  
   Eq.~(\ref{Laplacian.A.tor})
rewritten in terms of  $G^{\rm tor}$
with 
   Eq.~(\ref{radial.equation}) 
gives the main result  
of this section,
              Eq.~(\ref{main.result.radial.toroidal}).


In contrast to 
    \cite{Bicak}
we have given a 
     {\it simple, unified treatment} 
of all three cases, $K= ( 0, \pm 1 )$,
with 
     {\it form-identical derivations and results}.~---
Ref.~\cite{Bicak} 
discussed the solutions of the homogeneous momentum constraint
(which are eigenfunctions of the Laplacian) 
without referring to the Laplacian.


\section{MACH'S PRINCIPLE}
\label{Sec.Mach}


\subsection{Symmetries and Mach's principle}
\label{symmetries.Mach}



We treat 
    {\it all}
linear perturbation fields 
on
    {\it all}
types of FRW backgrounds, $K=(\pm 1, 0),$ 
and
    {\it all}
energy-momentum-stress tensors, 
i.e. 
    {\it all}
types of matter (also dark energy and a cosmological constant), 
not necessarily of the perfect-fluid form,
and for  
    {\it totally general field configurations}
of energy currents 
$J^{\varepsilon}_{\hat{k}} \equiv T^{\hat{0}}_{\, \, \hat{k}}.$
This is in striking contrast 
to the previous literature,  
which only treated 
the artificial situation of
spherical shells of matter rotating rigidly around one given axis.~---
However the 
    {\it mathematics} 
of totally general perturbations  
can be reduced to 
the mathematics of the special case
of spherical shells of matter in rigid rotation 
around a given gyroscope
with the help of 
      {\it two theorems,} 
which are based on the symmetries 
relevant for Mach's principle, rotation and parity:
%
\begin{enumerate}
\item  The 
         {\it precession} 
       of a 
         {\it gyroscope} 
       (relative to given local LONB axes)
         {\it cannot} 
       be caused by 
         {\it scalar}
       perturbations nor,
       in linear perturbation theory,  
       by 
        {\it tensor} 
       perturbations.~--- 
       In the vector sector
       the precession of a gyroscope can be caused only 
       by energy-current fields 
       with $J^P = 1^+$ 
       relative to the given gyroscope's position,
       i.e. of the 
         {\it toroidal}
       type and with $\ell =1.$
\item  On every mathematical spherical shell 
       centered on the gyroscope considered:
       The field component (in the harmonic decomposition 
       of the general energy-current field)
       which is toroidal and has $\ell = 1$
       (relative to the gyroscope) 
       is given by an  
           {\it equivalent rigid rotation}  
       with an 
           {\it equivalent angular velocity} of matter 
       $\vec{\Omega}^{\, \rm matter}_{\, \rm equiv} (\chi_{\rm s}).$
       The equivalent angular velocity of matter 
       is given by the 
            {\it global scalar product}
       of  the $\vec{J}_{\varepsilon}$-field 
       with the toroidal fields 
       $\vec{X}^{-}_{\ell =1,m}$ on the shell with radius $\chi:$
%
\begin{eqnarray}
&& \int d\Omega <\vec{X}^{- \, *}_{\ell =1,m} \, , 
\vec{J}_{\varepsilon} (\chi, \theta, \phi)>
\equiv  (J_{\varepsilon})^-_{\ell=1,m}(\chi)
\nonumber
\\ 
&& = -\sqrt{16 \pi/3} \, (\rho_0 + p_0) \, R(\chi) \,
[\Omega_m (\chi)]^{\rm matter}_{\rm equiv},
\label{projection}
\end{eqnarray}
where $<... \, , \, ...>$ denotes the point-wise inner product,
and $d\Omega$ is the element of solid angle,  
while $\Omega_m$ denotes spherical-basis components 
of the angular velocity.
The transformation between the spherical 
and the Cartesian components of $\vec{\Omega}$ is given by
$ \Omega_{m=0}     =  \Omega_z,$
and
$ \Omega_{m=\pm 1} = (\mp \Omega_x - i \Omega_y)/\sqrt{2}.$
\end{enumerate}
%

The proof of theorems (1) uses the following facts:
The precession of a gyroscope 
relative to given LONB axes, 
$d (S_{\hat{i}}) / dt,$ 
which equals the 
    {\it torque} 
on the gyroscope, 
is an 
     {\it axial vector,} 
$J^P = 1^+.$ 
The precession of a gyroscope is caused locally 
by the gravitomagnetic field $\vec{B}_{\rm g} (0),$
which is also an axial vector.~---
     For
       {\it scalar perturbations}
     all fields are derived from scalar fields (via differentiation),
     but this can only produce 
     source-fields $\vec{J}_{\varepsilon}$ 
     in the natural parity sequence, $0^+, 1^-, 2^+,$ etc,
     which cannot contribute to the precession.~---
     For
         {\it tensor perturbations}
     (gravitational waves)
     all linear perturbations are given by a
     traceless, divergenceless 3-tensor, 
     from which one cannot form an axial vector field
     to generate a $\vec{B}_{\rm g}$ field at the origin.


The proof of theorem (2) uses the following facts: 
For a general energy-momentum-stress tensor 
(not necessarily of the perfect-fluid type)
the local center-of-mass velocity $\vec{v}$ is given by 
$ v_{\hat{k}} \equiv  (\rho_0 + p_0)^{-1} T^{\hat{0}}_{\, \hat{k}}$ 
in linear perturbation theory, where $\rho_0$ and 
$p_0$ refer to the unperturbed FRW background.~---
For fixed $\chi,$ 
a toroidal velocity field $\vec{X}^-_{\ell=1,  m}$ 
is a flow equivalent to a rigid rotation of a shell of matter.


The gravitomagnetic 
       {\it vector potentials} 
$\vec{A}_{\rm g}$
relevant for the gyroscope's precession,  
must also be 
        {\it toroidal} 
with $\ell =1.$
The 
       {\it gravitomagnetic field} 
$\vec{B}_{\rm g} = {\rm curl} \vec{A}_{\rm g},$  
which causes the gyroscope's precession,        
must be 
    {\it poloidal} 
with $\ell=1.$


\subsection{Green functions for $(\Delta - \mu^2)$ on vorticity fields 
in open FRW universes}
\label{Green.open.FRW}


We derive the Green function
for the gravitational vector potential $\vec{A}_{\rm g}$
for $K = (-1,0)$  
generated by an energy-current source
which is toroidal, has $\ell =1, \, m =0,$
and is concentrated at  the geodesic distance $\chi_{\rm s}$
relative to the
gyroscope considered.
Such a source corresponds to 
a shell of cosmological matter in rigid rotation around the $z$-axis,
\begin{equation}
v^{\, \phi} = \Omega_{\rm s} \, \delta (\chi - \chi_{\rm s}).
\label{velocity.field} 
\end{equation}
%


The Green function for arbitrary $\ell$
is not needed to compute the dragging of inertial axes
at the origin.
The derivation for general $\ell$  
is totally analogous to the case $\ell = 1$
treated here,
and it gives the general solution 
of the momentum constraint 
for the toroidal sector.


As noted at the beginning of 
    Sec.~\ref{eigenfcts.Laplacian.scalar.fields}, 
all equations in 
    Secs.~\ref{eigenfcts.Laplacian.vorticity.fields} and \ref{Sec.Mach} 
involve the perturbation quantities at one time, 
and at the chosen time we set the scale factor 
$a$ equal to 
    {\it one.}


Inside and outside the rotating shell,
$\vec{A}$ is a 
     {\it toroidal vector eigenfield} 
of $(\Delta - \mu^2)$ according to the momentum constraint
   Eq.~(\ref{Einstein.equation.A.introduction}).
The wave number $q$ in the radial oscillations is
$q = \pm i \mu$ according to 
   Eq.~(\ref{main.result.radial.toroidal}).
We choose the positive sign.

      {\it Inside} 
the shell 
the radial function $G(\chi) = f(\chi)$ 
which multiplies $\vec{X}_{\ell =1, m}^-$ 
according to
      Eq.~(\ref{main.result.radial.toroidal})
is the generalized spherical Bessel function for 
$K = (\pm 1, 0), \, $
      Eq.~(\ref{radial.function.p-wave}),  
\begin{equation}
i \mu^2 \tilde{j}^{(K)}_{q=i\mu, \, \ell =1}(\chi)
=  \frac{-d}{d \chi}  
  [ \frac{1}{R} \,  \sinh(\mu \chi)].
\label{Bessel}
\end{equation}
Note that any {\it regular} toroidal vector field 
must {\it vanish} at the origin.

      {\it Outside} 
the shell 
the radial function for an 
     {\it open}
universe, $K=(-1, 0),$
which is 
     {\it regular} at spatial infinity for $q = i \mu$, 
is the generalized Hankel function of the first kind,
    Eq.~(\ref{Hankel.p-wave}), 
exponentially decreasing 
for $\chi \rightarrow \infty,$
\begin{eqnarray}
-i \mu^2 \, \tilde{h}^{(1)(K)}_{q=i\mu, \ell =1}(\chi)
&=& \frac{-d}{d\chi}[\frac{1}{R}\exp (- \mu \chi)]
\\
&=& Y_{\mu} (\chi) = \mbox{Yukawa force}. 
\label{Hankel}
\end{eqnarray}
Note that the Yukawa 
     {\it potential}
is $(1/R) \exp(-\mu \chi),$ 
while the Yukawa 
     {\it force} 
is minus the gradient (the $\chi$-derivative) 
of the potential.~---
For a closed universe this solution
is not acceptable, because the $1/R$ factor makes it 
     {\it singular} 
at $\chi = \pi,$ the 
     {\it antipodal point} 
to the origin.
This will be treated in   
     subsection~\ref{Green.closed.universe}.


The vector potential $\vec{A}^{\, {\rm tor}}$ 
is tangential to the shell, and 
it must be 
     {\it continuous across the shell},
because the component of $\vec{B}$ normal to the shell
must be continuous,
otherwise ${\mbox{div}} \vec{B} \neq 0$ on the shell.
In an open universe, 
$K= (-1, 0),$ we obtain
\begin{equation}
f^{{\rm tor}}_A (\chi, \chi_{\rm s}) 
= N \,
\tilde{j}       (\chi_{<}) \,
\tilde{h}^{(1)} (\chi_{>}),
\label{Green.radial.function}
\end{equation}
where we have dropped the subscripts $\{ q=i\mu, \, \ell =1 \}$
and the superscript $(K=-1, 0)$ for better readability.
The arguments $\chi_<$ resp. $\chi_>$ denote 
the smaller resp. larger of the two arguments $(\chi, \chi_{{\rm s}}),$
and $\chi_{{\rm s}}$ refers to the radius of the shell. 


The normalization factor $N$ is determined 
by the strength of the source $\vec{J}_{\epsilon}$ 
on the shell: 
Einstein's $G^{\hat{0}}_{\, \hat{\phi}}$ equation,
    Eq.~(\ref{Einstein.equation.A.introduction}),
determines the discontinuity of the tangential component
$B_{\hat{\theta}},$
\begin{eqnarray}
({\rm curl} &\vec{B}&)_{\hat{\phi}}^{{\rm sing}} 
=\delta(\chi-\chi_{\rm s}) \, {\rm disc}[B_{\hat{\theta}}]
=- 16\pi G_{\rm N} \, J^{\varepsilon}_{\hat{\phi}}
\nonumber
\\
&=& -\delta(\chi-\chi_{\rm s}) \, 16 \pi G_{\rm N} 
(\rho + p) \Omega_{\rm s} R_{\rm s} \sin \theta.
\label{disc.B.theta}
\end{eqnarray}
The discontinuity of $B_{\hat{\theta}}$
is determined by 
the discontinuity of its radial function
$[f^{({\rm pol, tg})}_B].$ 
For an open universe, $K = (-1, 0),$
    Eqs.~(\ref{div.pol.zero},
          \ref{curl.A.toroidal},
          \ref{Green.radial.function})
give
\begin{eqnarray}
&{\rm disc}& [f^{({\rm pol, tg})}_B]
= - \, {\rm disc}[\frac{d}{d \chi}f_A^{{\rm tor}}]
\nonumber
\\
&=& - N \, {\rm disc} [\, \frac{d}{d\chi} \{
\tilde{j}      (\chi_{<}) \, \,
\tilde{h}^{(1)}(\chi_{>}) \} \,]_{\rm s}
\nonumber
\\
&=&  - N \, W[\, 
\tilde{j}, \,
   \tilde{h}^{(1)} \, ]_{\rm s},
\label{disc.f.pol.tg}
\end{eqnarray}
where we denote the Wronskian by $W[f,g] \equiv (f g^{\prime} - f^{\prime} g)$
with $\prime = \frac{d}{d\chi}.$
Our Wronskian multiplied with $R^2$ is independent of $\chi,$
as can be seen from
    Eq.~(\ref{radial.equation}).
Therefore we evaluate $R^2W$ in the limit
$\chi \rightarrow 0.$ 
We use 
$\tilde{j}^{(K)}_{q=i\mu, \ell =1}(\chi) \rightarrow 
(i \chi \mu /3)(1 + K/ \mu^{2})$ and
$\tilde{n}^{(K)}_{q=i\mu, \ell =1} (\chi) \rightarrow
(\mu \chi)^{-2}$ for $\chi \ll 1,$
and we obtain
\begin{equation}
W[\, \tilde{j}_{q, \ell}^{(K)}, \, 
\tilde{h}^{(1)(K)}_{q, \ell} \, ]_{\, q=i \mu}^{\, \ell = 1} 
= \frac{1}{\mu R^2}(1 + \frac{K}{\mu^2})
\label{Wronskian}
\end{equation}
valid for $K = (\pm 1, 0)$ and for all $R.$
We note that for $(\ell =1, m=0)$
\begin{equation}
 (X^{+}_{\ell =1, m=0})_{\hat{\theta}} 
=(X^{-}_{\ell =1, m=0})_{\hat{\phi}} 
= - \sqrt{3/(4 \pi)} \sin \theta, 
\label{explicit.vector.harmonics}
\end{equation}
and all other components are zero.
We multiply ${\rm disc}[f_B^{({\rm pol, tg})}]$ with 
$(X^{+}_{\ell =1, m=0})_{\hat{\theta}}$
to obtain ${\rm disc}[B_{\hat{\theta}}],$
which we insert in
    Eq.~(\ref{disc.B.theta}) 
to determine $N,$
\begin{equation}
N = - (16 \pi G_{\rm N}) (\rho + p) R_{\rm s}^3 \Omega_{\rm s} 
\mu (1 + \frac{K}{\mu^2})^{-1} \sqrt{4 \pi/3}.
\label{N}
\end{equation}
This completes the computation of the Green function,
    Eq.~(\ref{Green.radial.function}),
in an open universe, $K = (-1, 0),$
for the vector potential generated 
by the 
toroidal current $\vec{J}_{\varepsilon}$
with $\ell =1, \, m=0$ at $\chi_{\rm s},$
specifically by the velocity field of
    Eq.~(\ref{velocity.field}).


The gravitomagnetic field $\vec{B}$ 
at the origin, the position of the gyroscope,
is obtained from  
    Eq.~(\ref{curl.A.toroidal}).
$\vec{B}_{\rm g} (0)$ written for arbitrary orientation 
of $\vec{\Omega}_{\rm s}$  
is
\begin{equation}
\vec{B} (0) = -2 \vec{\Omega}_{\rm gyro} (0) =
- \vec{\Omega}_{\rm s} \, 
\frac{2}{3} [16 \pi G_{\rm N} (\rho +p)] R_{\rm s}^3 \, 
Y_{\mu} (\chi_{\rm s}),              
\label{B.origin}
\end{equation}
valid in an open universe, $K =(-1, 0).$ 
This equation 
shows, 
how $\vec{\Omega}_{\rm s},$ 
  the angular velocity of the rotating shell,
determines $\vec{\Omega}_{{\rm gyro}},$ 
  the angular velocity of precession
  of a gyroscope at the center of the rotating shell.


In 
   Eq.~(\ref{B.origin})
we denote the prefactor by $\tilde{\mu}^2,$
\begin{eqnarray}
\tilde{\mu}^2 
&\equiv& 16 \pi G_{\rm N} (\rho +p)
\label{def.mu.tilde}
\\
\mu^2 
&\equiv& - 4 (dH/dt).
\label{def.mu.mu.tilde}
\end{eqnarray}
The connection between $\tilde{\mu}$ and
$\mu$ is given by
the $(dH/dt)$ equation,
which follows by subtracting the first from the second 
Friedmann equation,
\begin{eqnarray}
(dH/dt) - \frac{K}{a_{\rm c}^2} 
&=& - 4 \pi G_{\rm N} (\rho +p)
\nonumber
\\
a_{\rm c}  \equiv  1 \, \, \,
\Rightarrow \, \, \,
(\frac{\mu}{2})^2 + K &=& (\frac{ \tilde{\mu} }{2})^2.
\label{mu.tilde}
\end{eqnarray}
{\it Three length scales} occur in our problem:  
the $H$-dot radius $(\mu/2)^{-1},$  
the $(\rho +p)$-radius $(\tilde{\mu}/2)^{-1},$
and the curvature radius $a_{\rm c} \equiv 1.$


\subsection{Exact dragging of inertial axes for perturbations 
of an open FRW universe}
\label{Mach.open.universe}


The main result of this paper: 
(1)~The equations for ``Dragging of Inertial Axes'',
       Eqs.~(\ref{Mach.average.first.eq}, 
             \ref{Mach.average.introduction}), 
    are obtained directly from 
       Eq.~(\ref{B.origin}) 
    by integrating over all shell radii.
(2)~The equation, which shows that there is 
       {\it exact dragging} 
    (as opposed to partial dragging) 
    of local inertial axis directions,
    is the
    statement that the 
    weight function $W(\chi)$ 
    has its 
       {\it normalization} 
    equal to 
       {\it unity,}
       Eq.~(\ref{weight.normalization.introduction}).


We now give the proof 
that the normalization of the weight function
is indeed equal to 1 
for the case of an open universe, $K=(-1,0).$
First one does a partial integration 
of $(d/d\chi)$ in the Yukawa force.  
This gives
$W_{\rm tot} = 
\tilde{\mu}^2 \int d\chi R R^{\, \prime}  \exp (- \mu \chi).$
A partial integration of the factor $\exp (- \mu \chi)$ 
gives 
$W_{\rm tot} = 
\tilde{\mu}^2 \int d\chi (1 - 2 K R^2) \mu^{-1} \exp (- \mu \chi),$
where we have used
    Eq.~(\ref{R.double.prime}).
If instead of the last step, 
one performs a partial integration of the factor $(RR^{\prime})$,
one obtains
$W_{\rm tot} = 
\tilde{\mu}^2 \int d\chi (R^2/2) \mu \exp (- \mu \chi).$
Adding $\mu^2$  times the first expression 
plus   $(4K)$   times the second expression
one obtains
$(4K + \mu^2) W_{\rm tot} = \tilde{\mu}^2,$
and using
    Eq.~(\ref{mu.tilde})
gives $W_{\rm tot}^{(K=-1,0)} = 1.$


Our answer to Mach's question ``what share?'' is given by 
(1) the 
       {\it radial  weight function} 
    $W(\chi)$ with its Yukawa-force cutoff, and
(2) the 
       {\it angular weight function} 
    $\vec{X}^{- \, *}_{\ell=1,m}$ 
       of Eq.~(\ref{projection}) 
in the projection of the general velocity field 
on the toroidal vorticity sector with $\ell =1.$


The 
     {\it kinetic angular momentum} 
$\vec{L}$ per $d \chi$ is
directly obtained from $\vec{\Omega}_{\rm matter}$
using the
moment of inertia per $d\chi$ of a fluid shell, which is
$(8 \pi/3) (\rho + p) R^4,$ 
\begin{equation} 
\frac{d \vec{L}}{d \chi}
=   \frac{8 \pi}{3}  \, (\rho +p) \, R^{\, 4} \, 
\vec{\Omega}_{\rm matter} (\chi).
\label{moment.of.inertia.ang.momentum}  
\end{equation}
Evidently the  
     {\it kinetic} 
angular momentum $\vec{L}$ is determined by  angular velocity measurements
    {\it without prior knowledge}
of $g_{0 \phi},$ which is the output of solving Einstein's equations.


The  
     {\it gravito-magnetic moment} 
per $d\chi,$
$d \vec{\mu}_{{\rm g}} / d \chi,$
is defined in the same way as in ordinary magnetism, 
except that the current of charge is replaced 
by the current of energy  
$\vec{J}_{\varepsilon}:$
\begin{eqnarray}
\frac{d\vec{\mu}_{{\rm g}}}{d \chi} 
= \frac{1}{2} \frac{d \vec{L}}{d \chi}
&=&
\frac{1}{2} (4 \pi R^2) \int_0^{\infty} d\Omega \,
[\vec{R} \times \vec{J}_{\varepsilon}],
\label{magnetic.moment.vs.ang.momentum} 
\\
\int d\Omega \, [\vec{n} \times \vec{J}_{\varepsilon}]_m
&=& - \frac{4 \pi}{3} \int d\Omega 
<\vec{X}^{- \, *}_{\ell =1, m} \cdot \vec{J}_{\varepsilon}>,
\label{magnetic.moment.projection}
\end{eqnarray}
where $\vec{n}$ is the radial unit vector at the source point, 
and $\vec{R} = R \, \vec{n}.$
The last two equations show that 
the gravitomagnetic moment and the angular momentum
involve a projection of the energy-current field
on the sector of toroidal vorticity fields with $\ell =1.$


$\vec{\Omega}_{\rm gyro}$
is determined by an integral over the 
density of kinetic angular momentum:
From 
    Eqs.~(\ref{Mach.average.first.eq},
          \ref{moment.of.inertia.ang.momentum}) 
\begin{equation}
\vec{B}_{{\rm g}} (P) = -2 \vec{\Omega}_{\rm gyro} =
- 4G_{\rm N}
\int d({\rm vol}_Q) \,  \{ \frac{1}{R} 
\frac{d\vec{L}}{d({\rm vol})} \}
Y_{\mu} (\chi).
\label{B.in.terms.of.L}
\end{equation}
%


Expressing the kinetic angular momentum $\vec{L}$
by the energy current $\vec{J}_{\varepsilon}$
we obtain the 
  {\it fundamental law} 
for 
  {\it cosmological gravitomagnetism} 
in 
  {\it integral form} 
for an
  {\it open} 
FRW universe, 
$(K = -1, 0):$
\begin{eqnarray}
&& \vec{B}_{{\rm g}} (P) = -2 \vec{\Omega}_{\rm gyro} (P) =
\nonumber
\\
&& =-4G_{\rm N} \int d({\rm vol}_Q)  
[\vec{n}_{PQ} \times \vec{J}_{\varepsilon}(Q)] 
 Y_{\mu} (\chi_{PQ}).
\label{B.general}
\end{eqnarray}
Again:  
$\vec{\Omega}_{\rm gyro}$ is given by the sources  
$\vec{J}_{\varepsilon}$ at the 
    {\it same time.}
This follows directly from:
(1) The facts 
that dragging effects are given directly by the momentum constraint,
and that the momentum constraint in the vorticity sector,
      Eq.~(\ref{Einstein.equation.A.introduction}),
for the general
      {\it time-dependent} 
context of gravitomagnetodynamics is an 
    {\it elliptic equation} 
(it has no partial time-derivatives).
These facts have already been recognized by
   \cite{Lindblom.Brill, Lynden-Bell.1995, instantaneous.inertial.frame} 
among others.~---
(2) The fact that the tensor sector (gravitational waves) 
in linear perturbation theory
(and of course the scalar sector) 
cannot produce a gravitomagnetic field $\vec{B}_{\rm g}$ 
nor a torque on a gyroscope  has been demonstrated in 
      Sec.~\ref{symmetries.Mach}.


The 
    {\it vector structure} 
of 
    Eqs.~(\ref{magnetic.moment.vs.ang.momentum},
          \ref{magnetic.moment.projection},
          \ref{B.general}) 
needs explanation,
since there is no global parallelism 
on a hyperbolic 3-space $H^3.$ 
This is explained at the end of
   Sec.~\ref{introduction.solution}.


The differences between 
    Eq.~(\ref{B.general})
and 
    {\it Amp\`ere's law} in integral form are:
\begin{enumerate}
\item the replacement of the current of electric charge $J_q$
      by the energy current $J_{\varepsilon},$
\item the factor $(- G_{\rm N}),$ as in the transition from 
      Coulomb's law to Newton's law, 
\item the additional factor $4,$
      which occurs in the transition from Amp\`ere's law of 
      ordinary magnetism to gravitomagnetism, 
\item the replacement of the $1/r^2$ force
      in Amp\`ere's law 
      by the 
          {\it Yukawa force}  
      $Y_{\mu} (\chi)$ for 
          {\it cosmological} gravitomagnetism, 
\item the need to distinguish
      between $R$ (in the denominator of the Yukawa force) 
      and $\chi$ 
      (in the exponent and in the derivative) 
      for FRW backgrounds with 
          {\it curved} 
      3-space, $K = \pm 1,$
\item the remarkable fact that the 
          law~(\ref{B.general}) 
      for cosmological gravitomagnetism is
      valid for general 
          {\it non-stationary} situations,
      while Amp\`ere's law is valid only for 
      stationary currents. 
\end{enumerate}
%


What
     {\it distance interval} 
$\chi$ of sources 
     {\it dominates dragging} 
of local inertial frames at $P ?$~---
For a spatially flat universe, the weight function $W(\chi)$
grows linearly 
from $\chi = 0$ to $\chi \approx \mu^{-1} = \tilde{\mu}^{-1},$
and it decays exponentially 
for $\chi \gg \mu^{-1}.$
Hence the sources $Q$ around 
$\chi_{PQ} = \mu^{-1}$ 
dominate the frame-dragging at $P.$~---  
For a spatially hyperbolic universe, $K=-1,$ 
and for $\tilde{\mu} \gg a_{\rm c}^{-1} \equiv 1$
the situation is approximately the same as for $K=0,$
since the frame dragging is caused by sources 
at distances much smaller than the curvature radius.~--- 
But for $K = -1$ 
and for $\chi$ much larger than 
the curvature radius $a_{\rm c} \equiv 1,$
the effective exponential cutoff factor 
in the weight function $W(\chi)$ is
$\exp [-(\mu -2) \chi].$
From $(\rho + p) > 0$ and from 
   Eqs.~(\ref{def.mu.mu.tilde}, 
         \ref{mu.tilde})
with $K=-1,$
it follows that 
$\tilde{\mu} > 0 $ 
and    $\mu  > 2.$
For $\tilde{\mu} << 1$
we have $(\mu - 2) \approx \tilde{\mu}^2/4,$ 
the cutoff factor becomes
$\exp(- \tilde{\mu}^2 \chi/4),$
and we have three regimes for $W(\chi):$
For $0 \leq \chi << 1$   
the weight function $W(\chi)$ grows linearly,
for $1 << \chi << \tilde{\mu}^{-2}$ the weight function is constant,
and for $\chi >> \tilde{\mu}^{-2}$ it decays exponentially.
On a logarithmic scale the sources near 
$\chi = \tilde{\mu}^{-2}$ dominate frame dragging for
$\tilde{\mu} << 1.$~---
For all open universes
the dragging of inertial frames
is dominated by sources
at a distance equal or larger than the $(\rho + p)$-radius: 
Local inertial frames are rigidly
    {\it ``in the grip of the distant universe''.}


    {\it Form-invariance}
of the solution
    Eq.~(\ref{B.general})
under transformation to a rigidly rotating frame:
If the reference FIDO at the position of the gyroscope
changes his local spatial axes to new ones, $\vec{e}_{\hat{i}}^{\, *},$
with angular velocity  $-\vec{\Omega}^*(0)$ 
relative to the old ones,
    Eq.~(\ref{B.general})
remains valid as it stands,
because both sides of 
    Eq.~(\ref{B.general})
change by the same term.
On the right-hand-side
$(\vec{n} \times \vec{J}_{\varepsilon})$ changes by
$[-(\rho +p) \, \Omega^{\star} (0) \, 
R \, \sin \theta  \,  \vec{e}_{\hat{\theta}}],$
where we have put the $z$-axis of the spherical coordinates 
in the direction of $-\vec{\Omega}^{\star}.$
The integration over the solid angle gives
$(8 \pi  /3) (\rho + p) R \, \vec{\Omega}^{\star}(0).$
The integration over $\chi$ is the same as in 
     Eq.~(\ref{weight.normalization.introduction}).
The resulting change on the right-hand side is
$[-2 \vec{\Omega}^{\star}(0)],$
which is equal to the change on the left-hand side.


To obtain definite values for measurements
of fields $\vec{J}_{\varepsilon}$ and $\vec{B}_{\rm g},$
one must make an arbitrary choice of 
a definite state of rotation of FIDO-axis directions along 
     {\it one} 
world line,
as explained at the end of
    Sec.~\ref{Vorticity.Perturbations}. 
But for 
     {\it any} 
such choice: The field of 
      {\it measured energy currents} 
$\vec{J}_{\varepsilon}$ all by itself
      {\it completely determines} 
the   gravitomagnetic field $\vec{B}_{\rm g}$
at all points, i.e. the {\it precession} $\vec{\Omega}_{\rm gyro}$ 
of {\it gyroscopes} at all points:
    {\it No absolute element}
is needed in the input,
as required in the starting point of Mach
      \cite{Mach.Mechanik, Mach.Energy}.


For cosmological gravitomagnetism 
and for
     {\it given sources} 
$\vec{J}_{\varepsilon},$
measured via angular velocities, 
distances, and $(\rho + p)$:
       The solution 
         Eq.~(\ref{B.general})
       for the momentum constraint is 
         {\it unique.}
       There exists no ambiguity 
       of adding a solution of the homogeneous equation. 
         {\it No regular solution}
       exists 
       for the 
         {\it homogeneous} 
       momentum constraint 
       in cosmological gravitomagnetism, 
       $(\Delta - \mu^2) \vec{A}_{\rm g}=0.$~---    
       Therefore for regular solutions
          {\it no boundary conditions} 
       are needed at spatial infinity for open universes.
       See also Sect. XI of 
         \cite{CS.I}.


\subsection{Exact dragging of inertial axes 
for perturbations of a closed FRW universe}
\label{Green.closed.universe}


To find the Green function
for $\vec{A}_{\rm g}$ in a closed universe
and in the toroidal sector,
we note that a toroidal vector field 
$\vec{V}_{\ell =1, m}^{\rm tor}$ 
which is a 
     {\it regular vector field}
at $\chi = 0$ and at $\chi = \pi$
must have a point-wise norm which
     {\it vanishes}
linearly at these two points.
Therefore the radial solution 
$  G^{{\rm tor}}_A =  f^{{\rm tor}}_A,$ 
which multiplies $\vec{X}_{\ell =1, m}^-,$ 
must also vanish linearly at $\chi = 0$ and at $\chi = \pi.$
Hence the radial part of the Green function 
can be written as
\begin{equation}
f_A^{{\rm tor}} ( \chi, \chi_{\rm s})   
=
\bar{N} \,
  \tilde{j} (      \chi_{<}) \, 
  \tilde{j} (\pi - \chi_{>}).  
\label{Green.radial.function.k+1.Bessel}
\end{equation}
We have again dropped the subscripts $( q=i\mu, \, \ell =1 )$
and the superscript $(K=+1)$ for better readability.

The 
     {\it discontinuity} 
in the derivative of
the radial part of the Green function,
the analog of 
   Eq.~(\ref{disc.f.pol.tg}),
is
\begin{eqnarray}
{\rm disc} &[& \frac{d}{d \chi}f_A^{{\rm tor}} \, ]
\nonumber
\\
= \bar{N} \, W &[&\, \tilde{j} (\chi),  \,
                     \tilde{j} (\pi -\chi)  \, ]_{{\rm s}}.
\label{disc.f.pol.tg.k+1}
\end{eqnarray}
To evaluate the Wronskian $W$ we use
the method given before 
     Eq.~(\ref{Wronskian}).
Specifically, $(R^2 W)$ is independent of $\chi,$
and we evaluate this expression for $\chi \rightarrow 0.$
We use  
$\tilde{j}^{(K=+1)}_{q=i\mu, \ell =1}(\chi) \rightarrow 
(i/3) \chi \mu (1 + 1/ \mu^{2})$ 
and
$\tilde{j}^{(K=+1)}_{q=i\mu, \ell =1} (\pi - \chi) \rightarrow
i / (\chi^2 \mu^2) \sinh (\mu \pi)$ 
for $\chi \ll 1,$
and we obtain
\begin{eqnarray}
&& W[\, \tilde{j}_{q, \ell}^{(K)} (\chi) , \, 
\tilde{j}^{(K)}_{q, \ell} \, (\pi  - \chi)]_{\, q=i \mu, \, \ell =1}^{\, K=+1}
\nonumber
\\ 
&& = \frac{1}{\mu R^2} \, (1 + \frac{1}{\mu^2}) \, \sinh (\mu \pi).
\label{Wronskian.closed}
\end{eqnarray}
Hence the normalization $\bar{N}$ is
\begin{equation}
\bar{N} = N \sinh^{-1}(\mu \pi),
\label{N.bar}
\end{equation}
where $N \equiv N^{(K=0,-1)}$ is given by 
    Eq.~(\ref{N}).


The gravitomagnetic field $\vec{B}$ at the origin
generated by a rotating shell 
$\Omega (\chi) = \delta (\chi - \chi_{{\rm s}}) 
                         \Omega_{{\rm s}}$ 
on a FRW background with $K=+1$
is given by taking the 
        result~(\ref{B.origin})
for $K=(0,-1)$
and making the replacement
\begin{equation}
\exp(- \mu \chi) \, \Rightarrow \, 
\sinh^{-1}(\mu \pi) \,  \sinh[\mu(\pi - \chi)].
\label{replacement}
\end{equation}%
For an arbitrary velocity field 
it directly follows 
with this replacement
from the
       result of Eq.~(\ref{Mach.average.first.eq})
that  
$\vec{\Omega}_{\rm gyro}$ is equal to the weighted average
of $\vec{\Omega}_{\rm matter},$
\begin{eqnarray}
 \vec{\Omega}_{\rm gyro} 
&=& \int_0^\pi d\chi \, \,
\vec{\Omega}_{\rm matter}(\chi) \, \, W(\chi),
\nonumber   
\\
W(\chi) 
&=& \frac{1}{3} \tilde{\mu}^2 R^3 \sinh^{-1} (\mu \pi) \,
\nonumber
\\
&& \frac{-d}{d \chi} \{ \frac{1}{R} \sinh[\mu(\pi - \chi)] \}.
\label{Mach.average.closed}
\end{eqnarray}
%
Again, the averaging weight function is normalized to unity, 
$W_{\rm tot}= \int_0^{\pi} d \chi W(\chi) = 1.$
The proof uses the same steps as the ones given to prove
unit normalization of $W(\chi)$ for open universes 
in the second paragraph of
    Sec.~\ref{Mach.open.universe}.
Hence for perturbations of a closed FRW universe
the evolution of inertial axes 
     {\it exactly follows}
the 
     {\it weighted average} 
of cosmic matter motion,
there is 
     {\it exact dragging} 
of inertial axes
by cosmic energy currents, 
as stated in Mach's postulate.


The fundamental law for gravitomagnetism 
in integrated form 
follows with the 
   replacement~(\ref{replacement})
from
   Eq.~(\ref{B.general}),   
\begin{eqnarray}
\vec{B} (P) &=&  
- 4G_N 
\sinh^{-1}(\mu \pi )
\int d({\rm vol}_Q) \, 
(\vec{n}_{PQ} \times \vec{J}_{\varepsilon}) \, 
\nonumber
\\
&& \frac{-d}{d \chi} 
\{ \frac{1}{R} [ \sinh [\mu ( \pi  -  \chi  )]  \}. 
\label{B.general.k+1.new}
\end{eqnarray}
%


The 
    solutions~(\ref{Mach.average.closed}, 
               \ref{B.general.k+1.new})
are again
    {\it frame-invariant,} 
i.e. form-invariant 
when going to a reference frame
which is in globally rigid rotation 
relative to the previous reference frame.
The proof is identical to the one given for open universes in
    Sec.~\ref{Mach.open.universe}.


Einstein, Wheeler, and others, 
        recently e.g. Bi\v{c}\'ak et al
        \cite{Bicak},
have held the view
that from the Machian point of view closed universes
are preferable.
In contrast our results show that Mach's hypothesis,
exact dragging of inertial axes,
holds for linear perturbations of 
     {\it all} 
FRW universes, open or closed.


\subsection{Exact dragging of inertial axes 
for perturbations of Einstein's static, closed universe}


For a static, closed universe $\mu^2 \equiv -4 (dH/dt) =0,$
and $(\tilde{\mu}/2)^2 = K = +1.$
Taking the limit  $\mu \rightarrow 0$ of the result for
a FRW universe with $K=+1,$ 
   Eq.~(\ref{B.general.k+1.new}),
one obtains the fundamental law for gravitomagnetism 
for Einstein's static, closed universe,
\begin{eqnarray}
&& \vec{B} (P) =
\nonumber
\\  
&& - 4G_N 
\int d({\rm vol}_Q) \, 
(\vec{n}_{PQ} \times \vec{J}_{\varepsilon}) \,  
\frac{-d}{d \chi} [ \frac{1}{R} 
(1 - \chi / \pi) ] 
\label{B.general.Einstein}.
\end{eqnarray}
%
Again  
$\vec{\Omega}_{\rm gyro}$ is equal to the weighted average
of $\vec{\Omega}_{\rm matter},$
\begin{eqnarray}
 \vec{\Omega}_{\rm gyro} 
&=& \int_0^\pi d\chi \, \,
\vec{\Omega}_{\rm matter}(\chi) \, \, W(\chi),
\nonumber   
\\
W(\chi) 
&=& \frac{4}{3}  R^3          
\frac{-d}{d \chi} [ \frac{1}{R} (1 - \chi / \pi )],                    
\label{Mach.average.Einstein}
\end{eqnarray}
where $W_{\rm tot} = \int_0^{\pi} d \chi \, W(\chi) = 1.$
    Eq.~(\ref{Mach.average.Einstein}) 
states that also for perturbations of Einstein's static closed FRW universe
the evolution of inertial axes 
     {\it exactly follows}
the weighted average 
of cosmic matter motion,
there is 
     {\it exact dragging} 
of inertial axes
by cosmic energy currents, 
as stated in Mach's postulate.~---
The 
    solution~(\ref{B.general.Einstein})
is again 
    {\it form-invariant} 
when going to a reference frame
which is in globally rigid rotation 
relative to the previous reference frame.


Einstein's static, closed universe (3-sphere) 
has been important 
in the work of Ozsv\`ath and Sch\"ucking 
     \cite{Ozsvath}, 
who added a Bianchi IX amplitude (the lowest mode of tensor perturbations) 
to Einstein's universe,
and then investigated Mach's principle.
Unfortunately Ozsv\`ath and Sch\"ucking considered the vanishing of the 
    {\it local} 
vorticity relative to a gyroscope to be a test of Mach's principle.
This is in contradiction to our conclusion that the 
    {\it weighted cosmic average}
of the energy currents in
    Eq.~(\ref{Mach.average.Einstein})
directly determines the time-evolution of gyroscope axes.
Our result agrees with Mach's hypothesis that
some (to him unknown) cosmic average 
of matter motion
would determine the time-evolution of gyroscope axes.


\subsection{Mach's principle for the limits to the Milne 
and the de Sitter universe}
\label{Milne.deSitter}


Many physicists (but definitely not Mach) have asked again and again:
   ``If you could remove a body far away from all sources of gravity,
     what would be the inertial behaviour of this body?''~---
If the density of galaxies 
was e.g. $10^{-9}$ per Hubble volume 
in an otherwise empty universe,
a particle could be a distance of $10^3$ Hubble radii 
away from all other matter. 
All the same, if galaxies in the unperturbed universe 
were homogeneously and isotropically 
distributed, 
we have shown that for a universe with linear vorticity perturbations
    {\it Mach's principle still holds.}


The 
   {\it Milne universe} 
is the limit 
$\rho / \rho_{{\rm crit}} \rightarrow 0$ (for $p/\rho$ fixed)
of a FRW universe.
The first Friedmann equation gives
$H^2 + K/a_{\rm c}^2 \rightarrow 0,$
hence we have a hyperbolic universe
with a curvature radius $a_{\rm c}$
equal to the Hubble radius $H^{-1}.$ 
The Milne universe is equivalent to 
a part of Minkowski space-time, 
the forward light-cone originating at the space-time point
where the Hubble expansion 
(of test particles at rest 
in the Milne universe) started.

How can exact dragging of inertial axes 
by cosmic energy flows work  
for  $\rho / \rho_{{\rm crit}} \ll 1,$
when the total energy within a Hubble volume 
is arbitrarily small?
According to the discussion in
   Sec.~\ref{Mach.open.universe},
the weight function becomes $W(\chi) \approx
(\tilde{\mu}^2 /4) \, \exp(- \tilde{\mu}^2 \chi /4)$
for $\tilde{\mu}^2 
\equiv 16 \pi G_{\rm N} (\rho +p) \ll 1.$
The 
      {\it prefactor} 
$(\tilde{\mu}^2/4)$ 
is 
      {\it arbitrarily close to zero,}
but the integral for $W_{\rm tot}$
would be linearly divergent 
without the cutoff at 
$\chi_{\rm cutoff} = 4 \tilde{\mu}^{-2}.$
Therefore the 
     {\it integral without the prefactor} 
gives  $(4 / \tilde{\mu}^2),$
and this is 
          {\it arbitrarily large.} 
The product of these two factors gives $W_{\rm tot} = 1,$
exact dragging of inertial axes, as demonstrated in 
    Sec.~\ref{Mach.open.universe}.


The 
   {\it de Sitter universe} 
is the limit 
$(p/ \rho) \searrow (-1)$ 
of FRW universes with $K = 0, \pm 1.$
These three cases correspond to three different slicings 
of one de Sitter space-time geometry or part of it.


How does Mach's principle work
arbitrarily close to the de Sitter limit, 
i.e. for $[(p/ \rho) +1] \ll 1$ and 
$\rho / \rho_{{\rm crit}}$ finite and fixed,
when the energy currents within the Hubble volume
can be made arbitrarily small for any observer?


For $K=0$ we have $\mu^2 = \tilde{\mu}^2 = 16 \pi G (\rho + p).$
Hence the prefactor in $W(\chi)$
is of order $\mu^2 \ll 1.$
On the other hand the integral for $W_{\rm tot}$
would be quadratically divergent 
without the cutoff at $\chi_{\rm cutoff} = \mu^{-1}$.
Hence the integral without the prefactor is of order 
$1 / \mu^2 \gg 1.$
The two effects cancel, and $W_{\rm tot} = 1$
as stated in 
   Eq.~(\ref{weight.normalization.introduction}).


For $K=-1$
the discussion of the de Sitter limit,
$(\rho + p) \rightarrow 0,$
is exactly the same as the
discussion given above for the Milne limt.


For $K=+1$
and  $(\rho +p) \ll \rho,$
i.e.    $\tilde{\mu} \ll 1,$
one obtains $\mu \rightarrow \pm 2i (1-\tilde{\mu}^2/8),$
and $W(\chi) = 8 (3 \pi)^{-1} \sin^4\chi,$
i.e. $\tilde{\mu}^2$ drops out.
This gives
$\int_0^{\pi} W(\chi) d \chi= 1.$


\subsection{Globally rigid rotations: Zero modes of $(\Delta + 4K)$}
\label{Globally Rigid Rotations: Zero Modes}


We have seen in 
    Sec.~\ref{Vorticity.Perturbations} 
that for vorticity perturbations 
the fixed-time slices $\Sigma_t$ are uniquely determined, 
and that  each slice has the unperturbed 3-geometry 
of $\{ E^3, S^3, H^3  \}.$ 
For the unperturbed 3-geometries 
we have chosen unperturbed 3-coordinates:  
spherical coordinates for $\{ E^3, S^3, H^3  \},$
resp. Cartesian coordinates for $E^3.$
This does not yet fix the coordinatization (gauge) completely,
because there still exist 
   {\it residual gauge transformations}
by 
   {\it time-dependent globally rigid rotations} 
of the spherical coordinates for 3-space.
In this subsection 
we show that 
these residual gauge transformations are
     {\it zero modes} 
of $(\Delta + 4K),$
where $a_{\rm c}=1.$


Since this paper on Mach's principle 
focusses on rotational motion of fiducial axes,
not on linear acceleration of frames,
we apply a change of the spatially global (rigid) reference frame 
such that
the 3-coordinate velocity 
$dx^i/dt$ of the reference FIDO at the spatial origin, $\chi = 0,$
is left unchanged. 
The angular velocity of the 
      {\it new} 
reference-FIDO axes at the spatial origin
      {\it relative to the old} 
ones is denoted by
$[- \vec{\Omega}^{\star} (\chi = 0)].$


In the new coordinates  
the matter 3-velocity field is changed by the additional field 
$ \vec{v}^{\, \star}_{\rm matter}  $,
which is equal to the negative of  
the additional shift vector field $\vec{\beta}^{\star}$
and also equal to the negative of 
the additional gravitomagnetic vector potential $\vec{A}^{\star}_{\rm g}$, 
\begin{equation}
    \vec{v}^{\, \star}_{\rm matter}   
= - \vec{A}^{\, \star}_{\rm g}     
= \Omega^{\star}(0)  \, \vec{e}_{\phi}.  
\label{A.star}  
\end{equation}
The momentum constraint,
       Eq.~(\ref{Einstein.equation.A.introduction}), 
for the additional fields 
$  \vec{v}^{\, \star}_{\rm matter}  = - \vec{A}^{\, \star}_{\rm g},  $
with the definition of $\tilde{\mu}^2$  for the coefficient in the source term,
       Eq.~(\ref{def.mu.tilde}),
and with the  relation $ (\tilde{\mu}^2 - \mu^2) = 4 K,$
       Eq.~(\ref{mu.tilde}),  gives 
\begin{equation}
(\Delta + 4K)  \vec{A}^{\, \star}_{\rm g} = 0.
\label{zero.mode.equation}
\end{equation}
i.e. the additional field $\vec{A}_{\rm g}^{\, \star}$  
arising from a globally rigid rotation is a 
   {\it zero mode} 
of the operator $(\Delta + 4 K).$


The curl of  $\vec{A}^{\star}_{\rm g}$ gives the additional 
gravitomagnetic field  $\vec{B}^{\star}_{\rm g}$,
\begin{eqnarray}
B^{\star}_{\hat{\chi}}     &=& - 2 \Omega^{\star}(0) \cos \theta
\nonumber
\\
B^{\star}_{\hat{\theta}}  
&=& + 2 \Omega^{\star}(0) \sin \theta \, \frac{dR}{d \chi}
\nonumber
\\
B^{\star}_{\hat{\phi}}     &=& 0.
\label{B.field.hom.rot} 
\end{eqnarray}
For $K=0$ (Euclidean 3-space) 
the additional $\vec{B}^{\, \star}_{\rm g}$-field and
hence the $\vec{\Omega}^{\star}_{\rm gyro}$-field 
are homogeneous fields.
But for $K=\pm 1$ 
the additional $\vec{B}^{\, \star}_{\rm g}$-field is not a homogeneous field,
its magnitude varies as a function of $\chi.$


The additional fields 
$    \vec{v}^{\, \star}_{\rm matter}, \,  
       \vec{A}_{\rm g}^{\, \star}, \,
       \vec{B}_{\rm g}^{\, \star}$
are equal to the fields, which arise in an   
        {\it unperturbed FRW universe}, 
if the coordinate system is in rigid rotation relative to FRW coordinates, 
and if the FIDOs and their LONBs are adapted to the rotating coordinates.~---
But note that  the constraint for a given 
      {\it vanishing source field} $\vec{J}_{\varepsilon}\equiv 0$  
has the unique solution $\vec{A}_{\rm g} \equiv 0.$


\subsection{Six fundamental tests 
    \\ for the principle formulated by Mach}
\label{fundamental.tests}


The following six tests 
cover different aspects, intimately related, 
of the principle formulated by Mach:
%
\begin{enumerate}
\item        {\it Exact dragging}
         of local inertial axes by a weighted average of
         cosmological energy currents. 
         The 
             {\it weight function} 
         must be 
             {\it normalized to unity}.
\item    {\it ``Totally determined'':}
    The time-evolution of local inertial axes (relative to any given
    set of axes) 
    is totally determined
    by cosmological energy currents.
\item        {\it No absolute element as an input.} 
         The time-evolution of local inertial axes  
         must be an 
             {\it output} 
         determined
         by cosmic energy currents, 
             {\it not an input}.
\item  {\it ``Solutions Frame-Invariant'':} 
    The
        {\it solution}
    for $\vec{\Omega}_{\rm gyro}$
    in terms of cosmological energy currents
    must be
        {\it form-invariant}
    under transformations to globally rotating frames,
    i.e. {\it frame-invariant.}
\item   {\it No boundary conditions}
         at spatial infinity in open cosmologies
         are needed to totally determine 
         local non-rotating frames.
\item   {\it No reference-frame condition}
         along one world line
         is needed to totally determine 
         local non-rotating frames: 
         In open universes such a reference-frame condition 
         along a world line at spatial infinity is equivalent 
         to boundary conditions. 
         In closed universes such reference-frame
         conditions along one world line 
         play the analogous role to boundary conditions
         for open universes.
\end{enumerate}
%
Tests (1)-(4) represent different aspects of the postulate 
                        as formulated by Mach.~---
Test (3) is the postulate of the ``relativists''  
(Huyghens, Bishop Berkeley, Leibniz, Mach, and others)
that only 
     {\it relative motion} 
of physical objects is relevant.
This was the 
     {\it starting point} 
of Mach.~---
Test (4) is a necessary condition, 
formulated by Mach in his earliest writings
     \cite{Mach.Energy}
quoted in 
     Sec.~\ref{Conclusions}.~---
Test (1) is an 
    {\it explicit formulation of Mach's Principle,}
given in  
    Sec.~\ref{Conclusions},
while Tests (2)-(6) are less explicit about 
the implementation.~--- 
Tests (1) and (2) are unambiguous for 
    {\it linear} 
perturbation theory.
In the 
    {\it nonlinear} case, 
there is now agreement that gravitational waves make contributions.
Hence 
   Einstein's formulation of Mach's principle~\cite{Einstein},
   ``the (metric tensor) field $g_{\mu \nu}$ is 
   entirely determined ... by the energy tensor of matter''  
is untenable.
But the definition of ``energy currents'' ( = ``momentum densities'') 
and ``angular momentum densities'' 
of the gravitational field has been controversial.
Frame-dragging effects by 
``the angular momentum of gravitational waves'' 
has been investigated in special, non-cosmological models in   
    \cite{Bicak.2008.1,  Lynden-Bell.2008.2}.~---
Tests (3) and (4) are meaningful in the presence of gravitational waves,
even if the concept of energy currents of the gravitational field
is not available in a precise way.~---
Tests (5) and (6) are intimately related.


         The 
              {\it kinetic energy currents}, 
         i.e. the 
              {\it LONB components} 
         $T^{\hat{0}}_{\, \, \hat{k}}$ 
         can be 
              {\it measured without prior knowledge}
         of the local non-rotating frames. 
         Hence 
              {\it no absolute element} 
         is needed in the input for our approach.~---
         In contrast the canonical energy currents, 
         i.e. the coordinate-basis components $T^0_{\, \, k}$  
         cannot be measured without
         the prior knowledge of the local non-rotating frames
         all over the universe.
         Hence using coordinate-basis components $T^0_{\, \, k}$  
         means that one needs an absolute element as an input,
         e.g. in the approach emphasized and advocated by 
              \cite{Lynden-Bell.1995, Bicak, Bicak.2007}.


The six tests listed above are 
     {\it not fulfilled  explicitly} 
(manifestly), 
if one uses the momentum constraint 
for the coordinate-basis components $T^0_{\, \, k},$ 
as done and advocated e.g. in 
   \cite{Lynden-Bell.1995, Bicak, Bicak.2007}.
These papers give no proof 
that any one of the six tests is satisfied.


Einstein's theory of General Relativity applied to the 
     {\it solar system}
fails tests (1)-(5).
Hence it is not a 
``theory of relativity'' in the sense of the relativists before Einstein, 
as has been recognized for a long time.


In this paper we have shown that the six fundamental tests 
listed above are 
        {\it manifestly (explicitly) fulfilled} 
for linear perturbations of FRW universes in 
        {\it Cosmological General Relativity}.



\begin{acknowledgments}
We like to thank T.~Frankel, J.~Fr\"ohlich, G.M.~Graf, J.~Hartle, 
and N.~Straumann for discussions, 
and D.R.~Brill and R.M.~Wald for hospitality and discussions.
\end{acknowledgments}



\appendix*

\section{THE DE RHAM-HODGE LAPLACIAN FOR VECTOR FIELDS 
IN RIEMANNIAN 3-SPACES:
CONNECTING THE CALCULUS OF DIFFERENTIAL FORMS AND THE CURL NOTATION}
\label{appendix.diff.forms}


The material in this appendix (in addition to 
     Sec.~{\ref{subsection.Laplacian})  
is essential 
for cosmological gravitomagnetism, because:

(1) For vector perturbations the
           {\it de Rham-Hodge Laplacian} 
    (``the'' Laplacian in the mathematics literature) 
    has many crucial advantages (listed below) 
    over $\nabla^2,$ the ``rough Laplacian'',
    which, unfortunately, has been used in the entire literature on cosmology.
    The calculus of 
           {\it differential forms}
    is needed for the de Rham-Hodge Laplacian.
    In addition, this calculus makes  
           {\it computations simple,}
    since no Christoffel symbols are required for computations.
    Unfortunately this calculus has not been used 
    in the literature on cosmological vector perturbations.

(2) On the other hand the 
           {\it elementary notation}
    for vector calculus in Riemannian 3-spaces 
    with 
          {\it curl} 
    and 
          {\it div} 
    is indispensable 
    for 
         {\it physical insight}
    in gravitomagnetism,    i.e. for seeing directly that the 
    mathematical structure of the momentum constraint
    for cosmological gravitomagnetism 
    on FRW backgrounds with $K=(\pm 1, 0)$ is 
          {\it explicitly form-identical} 
    (without curvature terms)      
    to the familiar structure of
    Amp\`ere magnetism in Euclidean 3-space
    (except for the Yukawa cutoff at the $H$-dot radius).


(3) Unfortunately 
    almost all books on differential geometry
    and on General Relativity,
    and all research papers on cosmological perturbation theory
    do not treat the topics of the 
       {\it de Rham-Hodge Laplacian}
    and the 
       {\it Weitzenb\"ock}
    formula. 
The latter gives the difference 
    $(\Delta - \nabla^2) \vec{A}, \, $ 
    which is needed when comparing with the literature 
    on cosmological vector perturbations.
    In books about differential geometry these topics are treated  
    in 
         \cite{Jost, Frankel}.~---
    But, as far as we know, 
    no book on General Relativity or differential geometry
    and no paper on cosmological perturbations 
    gives the 
        {\it explicit equations} 
    which connect the notation of differential forms
    with the elementary notation 
    for vector calculus in Riemannian 3-spaces: 
    $(\mbox{curl} \, \vec{a})_{\mu} =  ( \star d \, \tilde{a})_{\mu},$
       Eq.~(\ref{curl}),
    and for vorticity fields
    $(\Delta \vec{a})_{\mu}  = - (\mbox{curl curl} \, \vec{a})_{\mu} 
    = - (\star \, d \star d \, \tilde{a})_{\mu},$
       Eqs.~(\ref{curl.curl},
             and      \ref{star.d.star.d}). 
    Ref.~\cite{Jost}
does not give such equations, and 
    Ref.~\cite{Frankel.p.94}
    instead gives a ``rough, symbolic identification'', 
    which cannot be used literally in the equation 
    $\Delta \vec{a} = - \mbox{curl curl} \, \vec{a},$   
    see the paragraph following the 
         identity~(\ref{curl}).


Our aim in this Appendix is to give 
the derivations of the tools and results 
needed in cosmological gravitomagnetism. 
We present this for cosmologists 
with little or no experience in the calculus of differential forms.


The identities for vector calculus in Riemann 3-spaces
at the level of 
    {\it first} 
covariant derivatives $\nabla$
are indistinguishable from 
identities for vector calculus  
in Euclidean $3$-space,
given e.g. in 
    \cite{Jackson.vector.formulas.top},
because 
    {\it curvature effects cannot appear} 
at the level of first derivatives.---
At the level of 
    {\it second}
covariant derivatives $\nabla,$ 
the standard identities of vector calculus
in Euclidean 3-space, e.g. in
    \cite{Jackson.vector.formulas.top},
remain true in Riemannian 3-space,
if and only if one uses 
the de Rham-Hodge Laplacian $\Delta$ as opposed to the 
`rough Laplacian' $\nabla^2.$


The reasons which 
       {\it single out} 
the de Rham-Hodge Laplacian in the vector sector on curved 3-space, 
and which give its 
       {\it conceptual motivation} 
in physics:
\begin{enumerate}
\item  If for a vector field all types of 
           {\it sources}
       (div and curl)  are 
           {\it zero,}  
       then the de Rham-Hodge Laplacian of this vector field 
       is also zero, i.e. the vector field is harmonic;
           Sec.~\ref{subsection.Laplacian}.
\item  The de Rham-Hodge Laplacian 
           {\it commutes} 
       with curl, div, grad;
           Sec.~\ref{subsection.Laplacian}.
\item  The 
           {\it identities} 
       of vector calculus in Euclidean 3-space
           \cite{Jackson.vector.formulas.top}
       and the important first identity of Green  
       generalized to vorticity fields,
            Eq.~(\ref{Greens.first.identity}),
       remain true in Riemannian 3-spaces,
       if and only if one uses 
       the de Rham-Hodge Laplacian.
\item  For electromagnetism in curved space-time the 
           {\it equivalence principle forbids curvature terms,}
       if and only if the de Rham-Hodge Laplacian 
       resp d'Alembertian is used;
           Sec.~\ref{subsection.Laplacian}.
\item  The 
           {\it action principle} 
       for Amp\`ere magnetism in Riemannian 3-space,
           Eq.~(\ref{action}), which is 
              {\it bilinear in first derivatives},     
       produces the Amp\`ere equation with the
       de Rham-Hodge Laplacian and without curvature terms, 
           Eq.~(\ref{curl.curl}).
\item  The momentum constraint for cosmological
           {\it gravitomagnetism} 
       has 
           {\it identical mathematical structure} 
       to 
           {\it Amp\`ere} 
       magnetism (except for the Yukawa cutoff), 
       if and only if 
       the de Rham-Hodge Laplacian
       is used;
           Eq.~(\ref{Einstein.equation.A.introduction}). 
\item   The 
                         {\it Hodge decomposition}
        for closed Riemannian 3-spaces, specialized to 1-forms, 
        states that any vector field can be uniquely decomposed 
        into a sum of a gradient field  plus a curl field  plus a
        harmonic field.
        This theorem is valid, if and only if the 
        de~Rham-Hodge Laplacian is used to define harmonic fields. 
\item      {\it No Christoffel symbols} 
        resp connection coeffiecients
        are needed to compute the de Rham-Hodge Laplacian;
           Eqs.~(\ref{curl},
                      \ref{div},
                      \ref{curl.curl.grad.div},
                      \ref{Laplace.diff.forms}).          
\end{enumerate}
Every one of these properties does 
   {\it not} 
hold for the ``rough Laplacian'' $\nabla^2.$


\subsection{Differential forms and curl, grad, div}


One can 
     {\it represent} 
a vector field $\vec{v}$ in Riemannian 3-space,
a geometric object, 
by its Local Ortho-Normal Basis (LONB) components, 
denoted by a 
     {\it hat} 
over the index, $v_{\hat{\ell}},$
or by its contravariant components in a coordinate basis
$v^{\lambda},$
or by its associated 1-form $\tilde{v},$ 
resp. by its 1-form components 
(covariant components) $v_{\lambda}.$
The calculus of differential forms is fundamentally
tied to the representation
in some unspecified {\it coordinate basis} 
and with {\it covariant} components.


For the
       {\it vector product}
$( \, \vec{a} \times \vec{b} \, )$ the two notations 
are connected by the equation
\begin{equation}
( \, \vec{a} \times \vec{b} \, )_{\lambda} 
= g_{\lambda \kappa} \varepsilon^{\kappa \mu \nu} a_{\mu}  b_{\nu}
= ( \, \star \, [ \, \tilde{a} \wedge \tilde{b} \, ] \, )_{\lambda},
\label{vector.product}
\end{equation}
where $\tilde{a}$ and $\tilde{b}$ are the 1-forms (covariant vectors)
associated with $\vec{a}$ and  $\vec{b},$ i.e.
$(\tilde{a})_{\mu} = (\vec{a})_{\mu} = a_{\mu}.$
A tilde $\tilde{}$ designates $p$-forms
(totally antisymmetric covariant tensors of rank $p$).~---
The first step is the
      {\it wedge product} 
$\wedge$ (exterior product, antisymmetric product) 
of two 1-forms, which gives a 2-form $\tilde{c}$
with components
$c_{\mu \nu} =   [ \, \tilde{a} \wedge \tilde{b} \, ]_{\mu \nu}
\equiv a_{\mu} b_{\nu}  - 
       a_{\nu} b_{\mu}.$~---
The second step is taking the 
     {\it Hodge dual} 
(Hodge star operator) of the resulting 2-form.
All our bases will have right-handed orientation.
The primary definition 
for the Hodge dual 
is given in any LONB: 
One contracts with the totally antisymmetric 
$\varepsilon$-tensor, the 
    {\it Levi-Civita tensor,}
in a LONB, where
$\varepsilon_{\hat{1} \hat{2} \hat{3}} \equiv +1.$
Explicitely: the Hodge dual of of a $p$-form $\tilde{c}$ is
given by 
$(\star \tilde{c})_{\hat{i}_1, .. , \hat{i}_{n-p}} =
(1/p!) \, \, \varepsilon_{\hat{i}_1, .. , \hat{i}_{n-p},
                          \hat{j}_1, .. , \hat{j}_{p}}
         \,   \,       c_{\hat{j}_1, .. , \hat{j}_{p}},$
where $n=3$ is the dimension of our Riemann space. 
Example: $\tilde{h}^{(1)} = \star \, \tilde{b}^{(2)}$ 
written in LONB components gives 
$h_{\hat{1}} = b_{\hat{2} \hat{3}}.$
It follows that $\star \star =1 $ 
for $p$-forms in a Riemannian 3-space.~---  
The 
    {\it volume element} 
spanned by a LONB 
with postive orientation is equal to +1:
$          v(\vec{e}_{\hat{1}}, 
             \vec{e}_{\hat{2}},
             \vec{e}_{\hat{3}}) =
  \varepsilon(\vec{e}_{\hat{1}}, 
             \vec{e}_{\hat{2}},
             \vec{e}_{\hat{3}}) =
\varepsilon_{\hat{1} \hat{2} \hat{3}}= +1.$
The LONB components of the $\varepsilon$-tensor 
are 
    {\it invariant under proper rotations}
(as are the LONB components of the metric tensor,
$g_{\hat{i} \hat{j}} = \delta_{ij}).$~---
But in the calculus of differential 
forms the Hodge dual
involves the totally antisymmetric $\varepsilon$-tensor 
in a 
    {\it coordinate basis,}
$\varepsilon_{\lambda \mu \nu} 
=\varepsilon_{\hat{\lambda} \hat{\mu} \hat{\nu}} 
\, \sqrt{g}$ and 
$\varepsilon^{\lambda \mu \nu} 
=\varepsilon_{\hat{\lambda} \hat{\mu} \hat{\nu}} 
\, (\sqrt{g})^{-1}.$
The volume of a parallelepiped spanned by a
triple of vectors is $v(X,Y,Z) = \varepsilon_{\lambda \mu \nu}
X^{\lambda} Y^{\mu} Z^{\nu},$
hence the volume 3-form $\tilde{v}$ has components
$v_{\lambda \mu \nu} =
\varepsilon_{\lambda \mu \nu}.$~---
  For the Hodge dual which takes a $p$-form into a $(3-p)$-form,
  one can either first    raise the indices 
  of the original $p$-form    and then act with 
  $\varepsilon_{\lambda \mu \nu},$ 
  or equivalently one can first    act with 
  $\varepsilon^{\lambda \mu \nu} $
  and then lower the indices to obtain
  the final $(3-p)$-form.


The operation 
       {\it curl} 
takes a vector field $\vec{a}$ 
into another vector field of opposite parity, 
$ (\, {\rm curl}  \, \vec{a} \, ) 
= ( \, \vec{\nabla} \times \vec{a} \, ), $  
where $\nabla$ is the covariant derivative.
The formula analogous to
       Eq.~(\ref{vector.product})
gives,   
\begin{equation} 
  ({\rm curl} \, \vec{a})_{\lambda} 
= (\vec{\nabla} \times \vec{a})_{\lambda} 
= g_{\lambda \kappa} \varepsilon^{\kappa \mu \nu} 
  \partial_{\mu}  a_{\nu}
= (\star \, d \, \tilde{a})_{\lambda}.
\label{curl}
\end{equation}
%
The last and the next-to-last expressions in 
     Equations~(\ref{curl})
give the 
    {\it simplest method}
to 
    {\it compute} a curl. 
But the first and second expressions are crucial to make the
    {\it precise connection} 
with the notation universally used in electrodynamics
     \cite{Jackson.vector.spherical.harmonics}
via an 
    {\it identity} 
in the form of an 
    {\it explicit equation}.~---
The first step on the right-hand side of
    Eq.~(\ref{curl}), the 
      {\it exterior derivative} 
(antisymmetric derivative) $d$
of a 1-form $\tilde{a}$ 
gives the $2$-form $d \tilde{a}$ with components
$(d \tilde{a})_{\mu \nu}
\equiv \partial_{\mu} a_{\nu}  - 
       \partial_{\nu} a_{\mu}.$
Because the calculus of forms is tied to 
(unspecified) coordinate bases
and to totally antisymmetric covariant tensors, 
the covariant derivative $\nabla_{\mu},$ 
after antisymmetrization,
can be replaced 
by the ordinary partial derivative $\partial_{\mu}.$
This is the 
      {\it first great simplification,}
and it takes place, 
because the connection coefficients in the coordinate basis, 
the Christoffel symbols 
$\Gamma^{\rho}_{\, \, \mu \nu},$
are symmetric in the second and third indices.~---
Compared to other identities given for curl in the literature 
      \cite{Frankel},
the identity 
$(\mbox{curl} \, \vec{a})_{\lambda} \equiv  (\star d \tilde{a})_{\lambda}$  
has the important advantage 
that it directly transcribes the computational prescription 
of elementary vector calculus for Cartesian coordinates,
$(\mbox{curl} \, \vec{a})_i = \varepsilon_{ijk} \partial_j a_k,$
to Riemannian 3-spaces 
(or to curvilinear coordinates in Euclidean 3-space).~---
Some authors, e.g.  
     \cite{Frankel.p.94},
make a 'dictionary' with the 'rough, symbolic identification'
${\rm curl \bf{A}} \Leftrightarrow  d \alpha^1,$
where the 1-form $\alpha^1$ is the covariant expression for 
the vector $\bf{A}.$ 
This identification cannot be used literally in the operation 
(curl curl $\bf{A}),$
because the Hodge star operator is missing in front of $d\alpha^1$
in their rough, symbolic identification.
The operation (curl curl) shows that
the output of the operation curl 
cannot be the 2-form $d\alpha^1,$
the output 
must be a vector 
(represented e.g. by a 1-form)
as in our 
   identity~(\ref{curl}),
which directly produces the identity
$(\mbox{curl curl} \, \vec{a} \, )_{\lambda} 
= (\star \, d \star d \, \tilde{a})_{\lambda}.$


The operation 
     {\it grad}
takes a scalar field $\phi$ into  
$\vec{\nabla} \phi$. 
%
\begin{equation}
( \mbox{grad} \, \phi)_{\lambda} = (\vec{\nabla} \phi)_{\lambda}=
\partial_{\lambda} \phi = (d \phi)_{\lambda}.
\label{grad}
\end{equation}
%


The 
      {\it inner product}
(scalar product) 
of two vectors
can be rewritten as an 
      {\it antisymmetric product}
by first going to the dual of $\tilde{b},$ 
\begin{equation}
\vec{a} \cdot \vec{b}
= a^{\lambda}  b_{\lambda}
= \frac{1}{2} \, \varepsilon_{\lambda \mu \nu} 
[ \, a^{\lambda} ( \varepsilon^{\mu \nu \kappa} b_{\kappa}) \, ]
= \star \, 
[\, \tilde{a} \, \wedge \, ( \, \star \, \tilde{b} \, ) \, ],
\label{scalar.product}
\end{equation}
where we have used 
$ \frac{1}{2}   \varepsilon_{\alpha   \beta  \gamma} 
                \varepsilon^{\beta    \gamma \delta} 
= \delta_{\alpha}^{\delta}.$
The wedge product 
of  a  1-form $\tilde{a}$ 
with a 2-form $\tilde{c}$ 
is given by 
$[\tilde{a} \wedge \tilde{c}]_{\lambda \mu \nu} 
                            = a_{\lambda} c_{\mu \nu} 
                            + a_{\mu} c_{\nu \lambda} 
                            + a_{\nu} c_{\lambda \mu}
=[\tilde{c} \wedge \tilde{a}]_{\lambda \mu \nu}.$ 
Similarly the inner product of two $p$-forms $\alpha, \beta$
is given by 
$<\alpha,\beta> = 
\star(\alpha\wedge\star\beta).$


The 
      {\it divergence,}
${\rm div} \, \vec{a} =  \, \vec{\nabla} \cdot \vec{a} \, ,$ 
can be written as an 
      {\it antisymmetric derivative}
in analogy with 
      Eq.~(\ref{scalar.product}).
This is the
      {\it second great simplification},
because the antisymmetrization again eliminates 
the need for Christoffel symbols.
It provides the easiest way 
to obtain the explicit expression for 
$( \,  \vec{\nabla} \cdot \vec{a}  \, )$ 
at the end of the following sequence of equations, 
\begin{eqnarray}
&& {\rm div}~\vec{a} = \vec{\nabla} \cdot \vec{a}  
= \star \, d \star \tilde{a} 
\nonumber
\\
&& = \varepsilon^{\alpha \beta \gamma} \partial_{\alpha}
  ( \varepsilon_{\beta \gamma \delta} a^{\delta} )
= (1/ \sqrt{g}) \, \partial_{\mu} \, ( \sqrt{g} \, a^{\mu}).
\label{div}
\end{eqnarray}
Some well-known textbooks on General Relativity 
need more than a dozen 
steps to derive the well-known last expression in 
     Eq.~(\ref{div}), because they work with Christoffel symbols.~---
But note that the expression $(\star \, d \star \tilde{a})$ 
is much more useful, since it 
gives e.g. $ {\rm div}~\vec{x}^{\pm}$ of 
    Eqs.~(\ref{div.x.plus},
          \ref{div.x.minus})
in a trivial way.


\subsection{The Laplacian on vector fields in curved 3-spaces}


The Laplacian $\Delta$ acting on vector fields resp 1-forms 
in curved space 
is called de Rham Laplacian
in  
     \cite{MTW.deRham},
and 
    \cite{Frankel.p.94}
writes: 
``This Laplacian was defined first 
by Kodaira and independently by Bidal and de Rham.''
The history including references is given in
     \cite{deRham.Var.Diff}, footnote on p. 125.
Since other authors use the name Hodge in this connection, and in order to be
clearly understood, we shall call it the de Rham-Hodge Laplacian.
The de Rham-Hodge Laplacian $\Delta$ must be distinguished 
from the 
     {\it ``rough Laplacian''}
$\nabla^2 \equiv \nabla^{\alpha}\nabla_{\alpha}$
    \cite{Berger},
    \cite{Frankel.p.94}.   
Only when applied to a scalar 
$\phi$ field, 
one has $(\Delta - \nabla^2) \phi = 0.$ --- 
For the primary conceptual definition 
of the Laplacian 
acting on vector fields, 
we start from an
     {\it action integral}, 
because an action integral only involves first derivatives,
hence it is independent of curvature effects.
For 
     {\it Amp\`ere magnetostatics} 
in static curved 3-space, 
the action integral 
(times $-1$)
reduces to the 
energy functional, 
\begin{equation}   
E[\vec{A}] = \int dv \, 
[ ({\rm curl}   \vec{A})^{2}
- 4 \pi \vec{A} \cdot \vec{J}\, ],
\label{action}   
\end{equation}   
where $ dv = d^3 x \, \sqrt{g}$ is the invariant volume element.
$E[\vec{A}]$ is given by the global scalar product  
of two vector fields,
$(.. \, , .. ) 
= \int dv \,
     \langle .. \, , .. \rangle,$
where $\langle .. \, , .. \rangle $ 
is the point-wise scalar product.


To carry out the variation   
of the energy functional $E[\vec{A}]$ 
with respect to variations of $\vec{A},$
one has to perform a partial integration.---
To make the surface terms disappear,
we must either 
have a compact Riemann space,
or, in a non-compact Riemann space,  
one of the two forms, the variation, must have 
   {\it compact support.}---
For partial integration in curved space
one needs the tool of differential forms. 
The 
      {\it adjoint operator} 
to $d$ in the sense of the global scalar
product
for compactly supported forms is denoted by~$d^*.$ 
Hence 
$( \beta^{\, p} , d \alpha^{\, p-1} ) \equiv  
( d^* \beta^{\, p}, \alpha^{\, p-1} ),$
where the superscript $p$  refers to $p$-forms.
Expressing the point-wise inner product via a wedge product,
      Eq.~(\ref{scalar.product}), 
and using the explicit coordinate-basis expression,
partial integration gives 
\begin{equation}
d^* \, \alpha^p = (-1)^p \, \star d  \star  \,  \alpha^p. 
\end{equation}
The superscript $^*$ denotes the adjoint,
the non-superscripted $\star$ denotes the Hodge dual.---
The 
    {\it codifferential operator} $\delta$
in this paper is defined with the opposite sign from the adjoint, 
\begin{equation}
\delta \equiv - d^* 
\, \, \, \Rightarrow \, \, \,
\delta \tilde{a} =  \mbox{div} \, \vec{a},
\end{equation}
where $\tilde{a}$ is a 1-form.
Some books define $\delta$ with the opposite sign from our sign,
$\delta \equiv d^*.$---
The exterior derivative $d$ 
   takes a $p$-form into a $(p+1)$-form,
the codifferential operator $\delta$ 
   takes a $p$-form into a $(p-1)$-form,
and $dd = 0, \, \delta \delta = 0.$


      {\it Green's first identity} 
generalized to vector fields with zero divergence
on Riemannian 3-spaces (partial integration) reads
\begin{equation}
\int dv \, \,
(\mbox{curl} \, \vec{\alpha}) \cdot (\mbox{curl} \, \vec{\beta})
=
-  \int dv \, \, \vec{\alpha} \cdot \Delta \vec{\beta} 
\label{Greens.first.identity}
\end{equation}
in the absence of surface terms. 
This identity is form-identical with the identity in Euclidean 3-spaces,
and it can be used to give the 
     {\it primary operational definition} 
of the 
     {\it Laplace operator} 
(de Rham-Hodge Laplacian)
on vector fields with zero divergence in Riemannian 3-spaces.~---
Equivalently, in our physics situation the primary concept
of the Laplacian is given by starting from the energy functional
     Eq.~(\ref{action}):
The factor multiplying 
the variation of $\vec{A}$ in the integrand of 
     Eq.~(\ref{action})
gives Amp\`ere's law for $\vec{A},$
i.e. $\Delta \vec{A}$ 
in the
    {\it vorticity sector}, 
\begin{eqnarray}
\Delta \vec{A} = - \mbox{curl curl} \vec{A} 
= - 4 \pi \vec{J}_{\rm q}
\label{curl.curl}
\\
(\Delta \vec{A})_{\mu} = - (\star \, d \star d \, \tilde{a})_{\mu}.
\label{star.d.star.d}
\end{eqnarray}
   Equation~(\ref{star.d.star.d})
with the operator $(\star \, d \star d \, )$
gives the simplest computational method 
in Riemannian 3-spaces 
(and for curvilinear coordinates in Euclidean 3-space).
On the other hand, 
   Eq.~(\ref{curl.curl})
with the operator (curl curl)
in Riemannian 3-spaces
is needed to make the precise connection 
using the notation universally adopted in electrodynamics
in Minkowski space in a (3+1)-split
     \cite{Jackson.vector.spherical.harmonics}.


In the general case of vector fields with
 div $\vec{a} \neq 0$ and
curl $\vec{a} \neq 0$
the requirements for the Laplace operator $\Delta$
of de Rham and Hodge 
acting on vector fields 
are:
\begin{eqnarray}
&& \mbox{(1) 
2nd order elliptic differential operator,} \, \, \, \, \, 
\, \, \, \, \, \, \, \, \, \, \, \, \, 
\nonumber
\\
&& \, \, \, \, \, \, \, \, \, \, \mbox{self-adjoint,}
\label{elliptic.selfadjoint}
\\ 
&& \mbox{(2) curl}~\vec{a} =0 
\, \, \, \mbox{\it and}  \, \, \, \mbox{div}~\vec{a} =0
 \, \, \, \,  \Rightarrow \, \, \, \, 
\Delta \vec{a} =0.
\label{Rightarrow}
\end{eqnarray}
%
Reversing the arrow in the 
     statement~(\ref{Rightarrow})
would be incorrect for 
    {\it open} 
Riemannian 3-spaces (open universes) 
in contrast to statements in some textbooks.
This is evident from simple counter-examples:
For a FRW background with $K=0$
a homogeneous $\vec{B}$-field gives 
$\Delta \vec{A}=0,$ but curl $\vec{A} \neq 0.$
For a FRW background with $K=-1$
the analogous field is a   
poloidal vorticity field $\vec{B}$
with $\ell=1$ and $q=0,$
for which again $\Delta \vec{A}=0,$ but curl $\vec{A} \neq 0.$~--- 
From
   Eqs.~(\ref{elliptic.selfadjoint},
         \ref{Rightarrow})
it follows that
\begin{eqnarray}
\Delta \vec{a} 
&=& - {\rm curl \, curl} \, \vec{a} 
+ {\rm grad \, div} \, \vec{a}. 
\label{curl.curl.grad.div}
\\
\Delta \tilde{a} 
&=& (- \star d \star d + d \star d \, \star) \, \tilde{a}
\nonumber
\\
&=& (\delta d + d \delta ) \, \tilde{a}
\nonumber
\\
&=& - (d + d^*)^2 \, \tilde{a}.
\label{Laplace.diff.forms}
\end{eqnarray}
%
Note that $(d d^*) $ is self-adjoint,
and $(d + d^*)^2 = (d d^* + d^* d ) $
is self-adjoint and positive semi-definite.


The 
    {\it ``rough Laplacian''} 
$\nabla^2 = g^{\mu \nu} \nabla_{\mu} \nabla_{\nu}$
does 
   {\it not} 
satisfy the requirement
$\{ {\rm curl}~\vec{a} =0$ and div~$\vec{a} =0 \}
\, \Rightarrow \, 
\Delta \vec{a} =0$ in curved space,
and it does 
   {\it not} 
satisfy Green's first identity 
   Eq.~(\ref{Greens.first.identity}).


\subsection{The Weitzenb\"ock Formula}
\label{section.Weitzenboeck}


The Weitzenb\"ock formula 
   \cite{Jost, Weitzenboeck}
for vector fields 
   \cite{Frankel}
and in 3-dimensional Riemannian space
is derived
by first using
(curl curl $A)_{\alpha} 
= - (\nabla^2 A)_{\alpha}
+ (\mbox{grad div} A)_{\alpha}
+~[(\nabla_{\beta} \nabla_{\alpha}
  - \nabla_{\alpha} \nabla_{\beta}) A]^{\beta} .$ 
For the last term one uses the Ricci identity,
\begin{equation}
[( \nabla_{\mu} \nabla_{\nu}    
- \nabla_{\nu} \nabla_{\mu} ) A]^{\alpha} 
= R^{\alpha}_{\, \, \beta \mu \nu}   A^{\beta},
\end{equation}
and the contracted Ricci identity,
\begin{equation}
[( \nabla_{\mu} \nabla_{\nu}    
- \nabla_{\nu} \nabla_{\mu} ) A]^{\mu} 
= R_{\nu \beta}   A^{\beta},
\end{equation}
where $ R^{\alpha}_{\, \, \beta \mu \nu}$ is the Riemann tensor and
$R_{\nu \beta}$ the Ricci tensor. 
The sign conventions for the Riemann and Ricci tensors 
are those of
Misner, Thorne, and Wheeler
    \cite{MTW}.
This gives
the difference between the true Laplacian 
$\Delta$ (de Rham-Hodge Laplacian) 
and the ``rough Laplacian'' 
$\nabla^{\alpha}\nabla_{\alpha},$
i.e. the Weitzenb\"ock formula,
\begin{equation}
(\Delta \tilde{A} 
- \nabla^2 \tilde{A})_{\alpha}=
R_{\alpha}^{\ \, \beta} A_{\beta}.   
\label{Weitzenboeck}
\end{equation}
For a FRW universe with $K = \pm 1$
and a curvature scale $a_c,$
the Ricci tensor of 3-space is $R_{\alpha}^{\, \, \beta} =  
(2K/a_{\rm c}^2) \,        \delta_{\alpha}^{\, \, \beta},$ 
and the Weitzenb\"ock formula gives 
\begin{equation} 
(\Delta - \nabla^{2})\tilde{A} 
= - (2K/a_{\rm c}^{2})   \tilde{A}.
\label{Weitzenboeck.FRW}
\end{equation}
%







\bibliography{paperY.bib}

\end{document}